\documentclass[fleqn,usenatbib]{mnras}
\usepackage{newtxtext,newtxmath}
\usepackage[T1]{fontenc}
\usepackage{upquote}
\usepackage{upgreek}
\usepackage{amsmath}
\usepackage{multicol}
\usepackage{array,multirow,graphicx}

\usepackage{tabularx}
\usepackage{hyperref}  
\usepackage[caption=false]{subfig}
\usepackage{float}
\usepackage{ulem}

\newcommand{\dm}{\rm DM}
\newcommand{\mdmunits}{{\rm pc \, cm^{-3}}} 
\newcommand{\dmunits}{$\mdmunits$}
\newcommand{\mdmstruct}{\dm_{\rm struct}}
\newcommand{\dmstruct}{$\mdmstruct$}
\def\code#1{\texttt{#1}}

\DeclareRobustCommand{\VAN}[3]{#2}
\let\VANthebibliography\thebibliography
\def\thebibliography{\DeclareRobustCommand{\VAN}[3]{##3}\VANthebibliography}

\title{An analysis of the time-frequency structure of several bursts from FRB\,121102 detected with MeerKAT}

\author[Platts et al.]{E. Platts,$^{1}$\thanks{E-mail: pltemm002@myuct.ac.za}
M. Caleb,$^{2}$
B. W. Stappers,$^{2}$
R. A. Main,$^{3}$
A. Weltman,$^{1,4}$
J. P. Shock,$^{1,4,5}$ \newauthor
M. Kramer,$^{3}$
M. C. Bezuidenhout,$^{2}$
F.~Jankowski,$^{2}$
V. Morello,$^{2}$
A. Possenti,$^{6,7}$ \newauthor
K. M. Rajwade,$^{2}$ 
L. Rhodes,$^{3,8}$
and J. Wu$^{3}$
\vspace{5mm}\\
$^{1}$High Energy Physics, Cosmology \& Astrophysics Theory (HEPCAT) Group, Department of Mathematics and Applied Mathematics, University of Cape Town, \\ Rondebosch 7701,
Cape Town, South Africa \\
$^{2}$Jodrell Bank Centre for Astrophysics, Department of Physics and Astronomy, The University of Manchester, Manchester, M13 9PL, UK \\
$^{3}$Max-Planck-Institut f\"{u}r Radioastronomie, Auf dem H\"{u}gel 69, D-53121 Bonn, Germany \\
$^{4}$National Institute for Theoretical Physics, Private Bag X1, Matieland, South Africa \\
$^{5}$The Laboratory for Quantum Gravity \& Strings, Department of Mathematics and Applied Mathematics, University of Cape Town, Rondebosch 7701, South Africa \\
$^{6}$INAF-Osservatorio Astronomico di Cagliari, via della Scienza 5, I-09047 Selargius (CA), Italy \\
 $^{7}$Universit\`{a} di Cagliari, Dipartimento di Fisica, S.P. Monserrato-Sestu Km 0.700, I-09042 Monserrato (CA), Italy\\
$^{8}$Astrophysics, Department of Physics, University of Oxford, Keble Road, Oxford OX1 3RH, UK 
}

\date{Accepted XXX. Received YYY; in original form ZZZ}

\pubyear{2021}

\begin{document}
\label{firstpage}
\pagerange{\pageref{firstpage}--\pageref{lastpage}}
\maketitle

\begin{abstract}
We present a detailed study of the complex time-frequency structure of a sample of previously reported bursts of FRB\,121102 detected with the MeerKAT telescope in September 2019. The wide contiguous bandwidth of these observations have revealed a complex bifurcating structure in some bursts at $1250$\,MHz. When de-dispersed to their structure-optimised dispersion measures, two of the bursts show a clear deviation from the cold plasma dispersion relationship below $1250$\,MHz. We find a differential dispersion measure of ${\sim}1{-}2~\mdmunits$ between the lower and higher frequency regions of each burst. We investigate the possibility of plasma lensing by Gaussian lenses of ${\sim}10$\,AU in the host galaxy, and demonstrate that they can qualitatively produce some of the observed burst morphologies. Other possible causes for the observed frequency dependence, such as Faraday delay, are also discussed. Unresolved sub-components in the bursts, however, may have led to an incorrect DM determination. We hence advise exercising caution when considering bursts in isolation. We analyse the presence of two apparent burst pairs. One of these pairs is a potential example of upward frequency drift. The possibility that burst pairs are echoes is also discussed. The average structure-optimised dispersion measure is found to be $563.5\pm 0.2 (\text{sys}) \pm 0.8 (\text{stat})\,\mdmunits$ -- consistent with the values reported in 2018. We use two independent methods to determine the structure-optimised dispersion measure of the bursts: the \code{DM\_phase} algorithm and autocorrelation functions. The latter -- originally developed for pulsar analysis -- is applied to FRBs for the first time in this paper.
\end{abstract}


%%%%%%%%%%%%%%%%%%%%%%%%%%%%%%%%%%%%%%%%%%%%%%%%%%

\section{Introduction}
Discovered just over a decade ago \citep{Lorimer:2007qn}, fast radio bursts (FRBs) are one of the newest astrophysical enigmas. Despite a limited number of detections (${\sim}140$ published sources in the Transient Name Server (TNS)),\footnote{Available at the \url{https://www.wis-tns.org/}.} great strides have recently been made in narrowing down likely progenitors. Earlier this year, for example, an FRB-like event was associated with a Galactic magnetar \citep[FRB\,200428;][]{Bochenek_2020,Andersen_2020}. However, due to the extensive range in energetics and activity levels of FRBs, not all can be attributed to a Milky Way-like population of magnetars \citep{2017ApJ...843L..26B,2019ApJ...886..110M,2020ApJ...893....9Z,2020Natur.577..190M,2020ApJ...899L..27M}. One such example is FRB\,121102 \citep{Spitler:2014fla,Spitler:2016dmz}, whose prolific repetitions have made it one of the most well-studied FRBs to date. Targeted multi-wavelength campaigns have revealed coincident persistent radio and optical emission \citep{Chatterjee_2017,Marcote_2017}. Using spectroscopic data from the optical source, \citet{Tendulkar_2017} calculated the redshift to be $z=0.19273(8)$. FRB\,121102 thus became the first FRB to be localised to a host galaxy: a low-metallicity dwarf. This lead many to consider a possible connection between FRBs and young magnetars born in rare superluminous supernovae events \citep[SLSNe Type I; e.g. ][]{2019ApJ...886..110M}. High resolution optical imaging was then used to pin-point the FRB to a star-forming region in the galaxy \citep{2017ApJ...843L...8B,2017ApJ...844...95K}. As well as being well-localised, FRB\,121102 goes through active phases, with a possible period of ${\sim}157\,$days \citep{rajwade+20,2020arXiv200803461C}. This has further facilitated targeted observing campaigns.

Polarisation measurements have revealed the extreme and dynamic magneto-ionic environment of FRB\,121102: emission was found to be nearly 100\% linearly polarised with a rotation measure (RM) of $1.46\times10^5$\,rad\,m$^{-2}$ that decreased to $1.33\times10^5$\,rad\,m$^{-2}$ over a 7 month period \citep{Michilli_2018}. This rapid change in RM without a comparable change in the dispersion measure (DM) implies extreme variation in the line-of-sight projected magnetic field. As noted by \citet{Cordes_2019}, such large variation has only been seen near the Galactic center magnetar J1745$-$2900 \citep{2018ApJ...852L..12D}.

The source of the persistent radio emission is currently unknown. It may be from a weak active galactic nucleus \citep[AGN; e.g.][]{Marcote_2017} or from a magnetised electron-ion nebula \citep[][]{2019MNRAS.485.4091M}. Despite numerous follow-up searches, no prompt optical, X-ray or gamma-ray counterparts have been detected \citep[e.g.][]{2016ApJ...833..177S,2017ApJ...846...80S,2017MNRAS.472.2800H,2018MNRAS.481.2479M}. FRB\,121102 has been observed over a broad range of radio frequencies: from 600\,MHz \citep{Josephy_2019,2020MNRAS.496.4565C} to 8\,GHz \citep{2017ApJ...850...76L,2018ApJ...863....2G,2018ApJ...863..150S}. This has revealed a wide variety of time-frequency structures \citep[e.g.][]{Hessels2019}.

A common feature of repeating FRBs is a downward drift in frequency, where sub-bursts that arrive at later times have lower central frequencies \citep[e.g.][]{Hessels2019,andersen_2019,Amiri_2020,Fonseca_2020}. Higher frequency sub-bursts also appear to have shorter temporal durations \citep{2018ApJ...863....2G,Hessels2019,Amiri_2019b,Josephy_2019}. Further, it has recently been shown that three repeating FRBs (FRB\,121102, FRB\,180916.J0158+65 and FRB\,180814.J0422+73) have an inverse relationship between the frequency drift rate and temporal durations of sub-bursts \citep{2020arXiv201014041C}.  A number of models have been proposed to explain these phenomena, invoking intrinsic mechanisms, propagation effects, or a combination thereof \citep[e.g.][]{Hessels2019}. Intrinsic mechanisms include pulsar-like sparking and cosmic-comb models \citep{Wang_2019,Wang_2020}, radius-to-frequency mapping in pulsars \citep{Lyutikov_2019}, the decreasing Lorentz factor of electrons near the surface of a neutron star \citep{2020MNRAS.497.1543G}, decelerating blastwaves from the flare ejecta of young magnetars \citep{2019MNRAS.485.4091M}, or the (potentially relativistic) motion of highly collimated FRB emission with respect to an observer \citep{2020MNRAS.498.4936R}. Propagation effects include scintillation \citep{simard+18,2020ApJ...899L..21S} and plasma lensing \citep{Cordes_2017}. \citet{2020ApJ...899L..21S}, however, find scintillation to be inconsistent with the measured drift rate of FRB\,121102. Further, one would expect to observe upward and downward drift in roughly equal parts. Plasma lensing bares a similar shortfall: the lack of upward drift reported in repeating FRBs requires a (rather unlikely) single dominant lens.
 
A definite example of upward drifting has yet to be reported, however it might be present in some pairs of closely-separated FRB bursts. Here, the second burst arrives at a higher frequency than the first, for example in FRB\,180916.J0158+65 \citep{Chawla_2020,Amiri_2020}, FRB\,190611 \citep{2020MNRAS.497.3335D}, FRB\,200428 \citep{Bochenek_2020,Andersen_2020}, and burst 03 in \citet{2020MNRAS.496.4565C}. In these cases it is unclear whether the sub-bursts are indeed emitted within the same burst envelope or are independent. \citet{2020ApJ...899L..21S} show that the first two bursts of FRB\,200428 were likely emitted within the same burst envelope and that the observed drift may be a result of scintillation. In the case of burst 03, however, there is no discernible scintillation. The upward drift may evidence lensing, but it is unclear whether the bursts are indeed from the same event.

Since the burst morphology of FRB\,121102, and some other repeating FRBs \citep[e.g.][]{andersen_2019,Fonseca_2020,2020MNRAS.497.3335D}, evolves with frequency, there is ambiguity between burst structure and the DM \citep[e.g.][]{2018ApJ...863....2G}. The emission of sub-bursts with different intrinsic central frequencies close in time, as well as propagation effects, can complicate accurate DM determination. For example, a burst that appears to have a different DM to other bursts may be made up of unresolved sub-bursts that drift down in frequency \citep[e.g.][]{2020arXiv200714404M}. To understand the mechanisms driving FRBs, it is essential that features intrinsic to FRBs are resolved. In maximising the frequency-averaged burst structure, one can determine sub-burst timescales and calculate frequency drift rates.

In Paper I \citep{2020MNRAS.496.4565C}, we presented 11 detections of FRB\,121102 made using the MeerTRAP system \citep{2018IAUS..337..406S} and single burst detection pipeline at the MeerKAT radio telescope \citep{2016mks..confE...1J}. Observations were taken over a ${\sim}3$ hour period on the 10th of September 2019 during the active phase of the FRB. Some of these bursts were observed to have complex frequency structure similar to those seen by \citet{Hessels2019}, with a few showing downward drifting substructure. MeerKAT's wide band receiver ($900{-}1670$\,MHz usable L-band range) allowed a detailed analysis of this complex frequency structure and frequency-dependant sub-burst drifting at a relatively low frequency. A number of intriguing features were noted, one of which is an apparent change in behaviour of some bursts at frequencies around $1250$\,MHz. Here, emission either became significantly fainter, exhibited a complex bifurcated substructure or appeared to deviate from the expected frequency-dependant arrival time ($\Updelta t\sim \nu^{-2}$). Two of the bursts (bursts 03 and 05) were each observed with a small `precursor' separated from the main burst by ${\sim}28$\,ms and ${\sim}34$\,ms, respectively, with the signal level between bursts equal to the noise floor. Three bursts had observable sub-bursts and the remaining six bursts had no discernible underlying structure.

In this paper we provide the structure-optimised DMs of the bursts presented in \citet{2020MNRAS.496.4565C} and give an analysis of the observed time-frequency structures. The paper is organised as follows: in Section \ref{sec:datared} we briefly detail the data reduction and in Section \ref{sec:method} we present the two algorithms used to calculate the structure-optimised DM -- Auto-Correlation Functions \citep[ACFs; e.g.][]{Cordes1990, Lange1998} and \code{DM\_phase} \citep{Seymour_2019}. Section \ref{sec:results_discuss} provides the results and a discussion of the observed burst features, and Section \ref{sec:conclude} concludes the paper.

%%%%%%%%%%%%%%%%%%%%%%%%%%%%%%%%%%%%%%%%%%%%%%%%%%
\section{Data reduction}
\label{sec:datared}
A detailed description of the data capture is given in Paper I  \citep{2020MNRAS.496.4565C}. The data contain only total intensity information (Stokes I only), with 4096 channels over a 856\,MHz bandwidth, and a time-resolution of $306\,\upmu$s, centered on 1284\,MHz. We cleaned the data manually for each burst using \code{pazi} in PSRCHIVE\footnote{Available \url{http://psrchive.sourceforge.net}.} \citep{2012AR&T....9..237V} to remove corrupt frequency channels. This masked a total of ${\sim}30\%$ of the band.

\section{DM Determination}
\label{sec:method}
\subsection{Maximizing burst structure}

\citet{Hessels2019} argued that a DM metric in which the frequency-averaged burst structure is maximised is more appropriate than maximizing the peak signal-to-noise (S/N). This structure-optimised DM corresponds to the DM value at which each sub-burst is correctly de-dispersed. \citet{Hessels2019} calculate the optimal DMs by maximizing the steepness of peaks in the frequency-averaged profile; specifically, they find the DM that maximises the mean square of each profile's forward difference time derivative \citep[also see ][]{2018ApJ...863....2G}. The relatively low time-resolution of the MeerKAT data ($306.24\,\upmu\rm s$), necessitated a different approach; although we note the methods used here are applicable to high resolution data, too. We implement two different techniques. In the first, Auto-Correlation Functions (ACFs) are used to determine the widths of structures in each burst \citep[e.g.][]{Cordes1990, Lange1998}. Here, the structure-optimised DM is that which minimises the widths. The second invokes \code{DM\_phase},\footnote{Available \url{https://github.com/danielemichilli/DM_phase}.} where the structure-optimised DM is found by maximising the coherent power across the bandwidth \citep{Seymour_2019}. For the analysis, the data were de-dispersed over a trial range of $540.0 \leq \dm \leq 590.0 \, \mdmunits$ with steps of $0.1 \, \mdmunits$. This step size was found to be suitable for bursts whose morphology evolved significantly with DM. This was not the case for all bursts -- for example, see the top panel of Figure \ref{fig:acf_and_burst}. Here, there is an unchanging time lag for numerous consecutive DM values. We chose not to increase the step size in such instances as it did not affect the results significantly.

\subsection{Autocorrelation functions}
ACFs give the correlation of a signal with a delayed copy of itself over different delay times $\tau=t_2-t_1$. They are a useful tool in determining microstructure time-scales of pulsar signals \citep[e.g.][]{Hankins_1972, Cordes1990,Lange1998}, and prove to be appropriate for our analysis of FRB sub-structure. The frequency-averaged ACF of a single burst, $f(t)$, is given by
\begin{equation}
    \label{eq:acf}
    \text{ACF}(\tau) = \int_{-\infty}^\infty f(t)\overline{f(t-\tau)} \text{d}t \;\;,
\end{equation}
where $\tau$ is the time lag and $\overline{f(t)}$ denotes the complex conjugate of $f(t)$. The narrow structures of the burst contribute to the ACF up to a scale that corresponds to their burst width ($\upDelta t_s$). As such, the presence of narrow structure is evidenced by a flattening in the ACF, i.e. where the ACF flattens, the narrow features no longer contribute to it. The lower the time lag value at which the ACF flattens, the shorter the burst width of these narrow structures and the more enhanced the sub-structure.

\begin{figure}

\subfloat{
	\includegraphics[width=\linewidth]{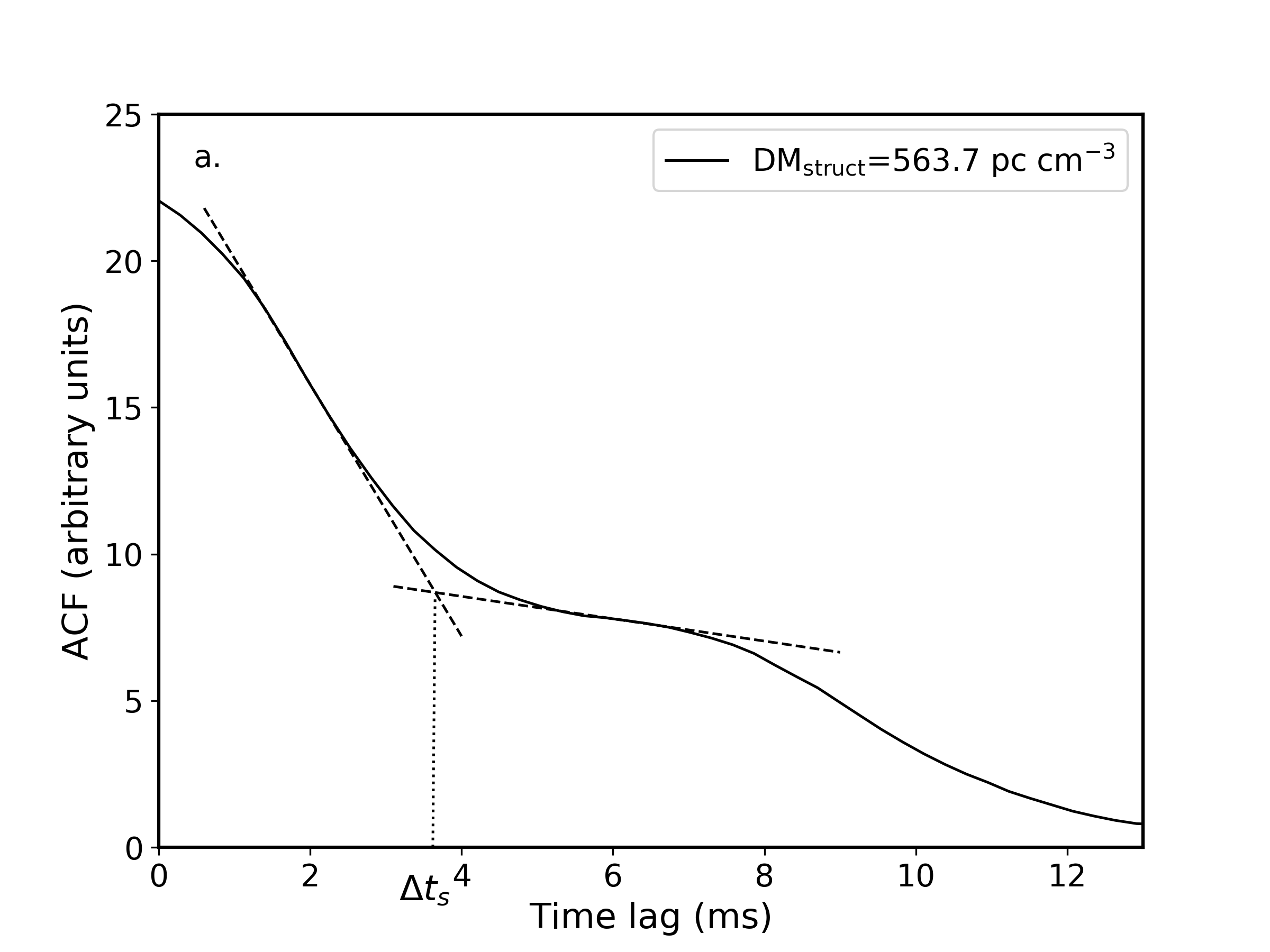}
	\label{fig:acf}}

\subfloat{
	\includegraphics[width=\linewidth]{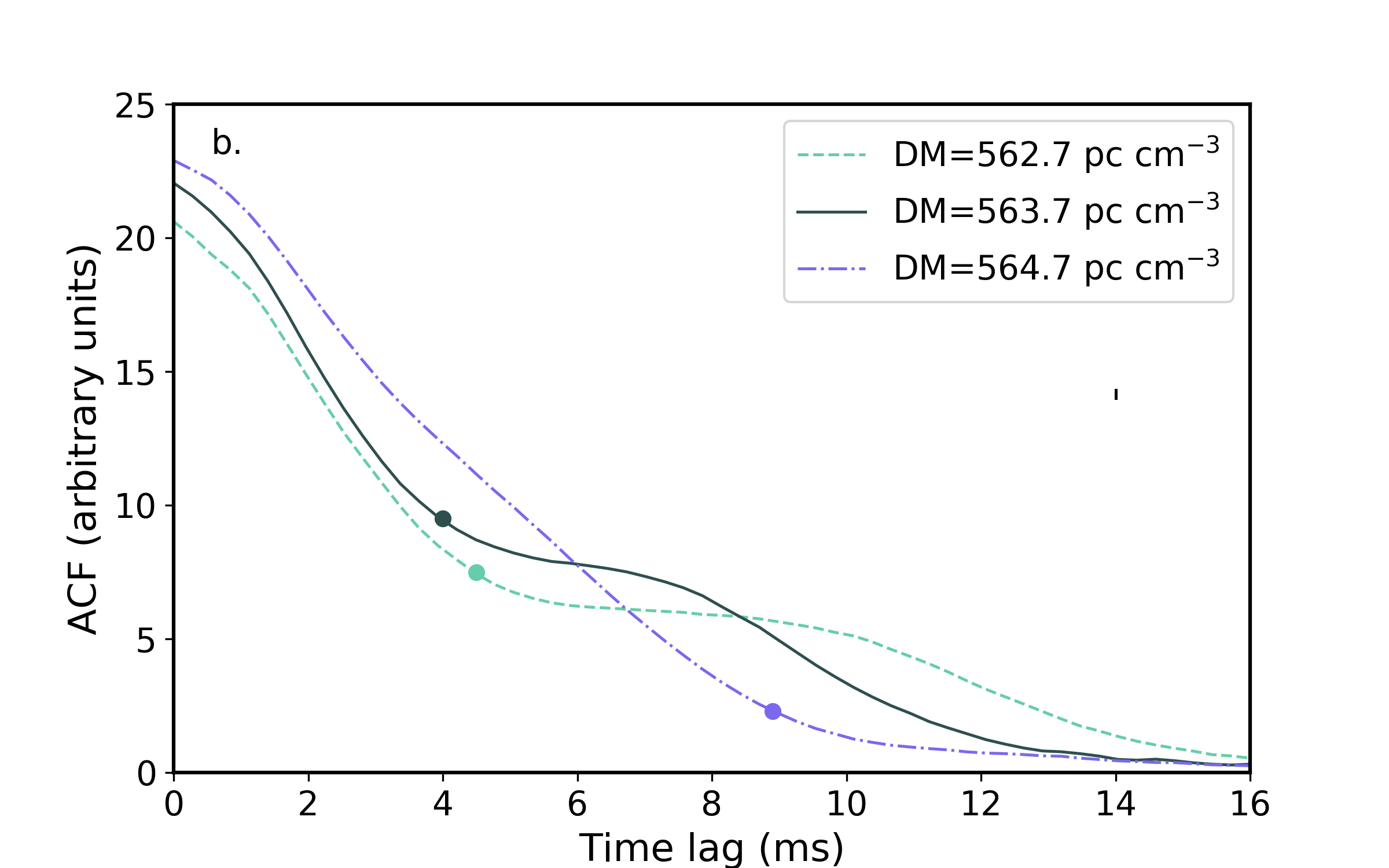}
	\label{fig:acf_dms}}

\caption{a. A schematic of the frequency-averaged ACF for burst 11 with $\mdmstruct=563.7 \, \mdmunits$. The sub-burst is depicted by the first bump, whose structure contributes up to a time-scale of $\upDelta t_s$ ms -- the point at which the ACF first flattens. The tangents are fitted by eye to illustrate the concept. b. An example of ACFs for burst 11 de-dispersed to different DMs. The circles correspond to the points of flattening given by the TOP algorithm} and give the structure time-scales. At the structure-optimised DM of $563.7\,\mdmunits$, the time lag is minimised. Note that by $\dm=564.7\,\mdmunits$, the ACF smooths out, driving the flattening point to much lower ACF values.
\end{figure}

\citet{Hankins_1972} defines the point at which an ACF flattens as the point of intersection of tangents fitted to the first (narrow) ACF region and the following (broad) ACF region (Figure \ref{fig:acf}). In a bid to automate this process, \citet{Lange1998} developed the Turn-Off Point (TOP) algorithm. Here, instead of fitting tangents by hand, a point of `significant flattening' is located by comparing the gradient of the ACF in different regions.

\begin{figure}
\subfloat{
	\includegraphics[width=\linewidth]{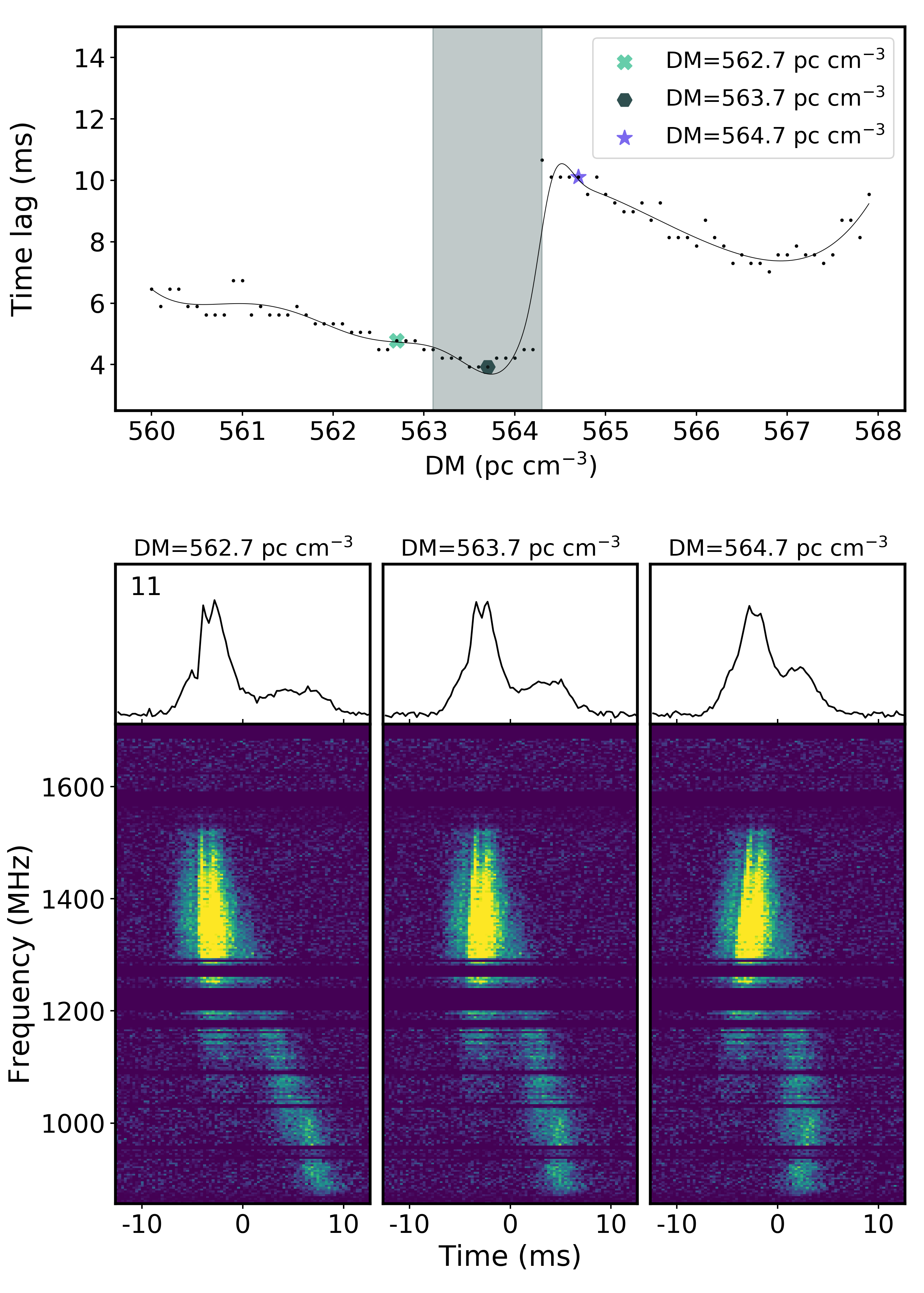}
	}
\caption{The top panel shows the DM vs time lag ($\Updelta t_s$) for burst 11. The shaded region corresponds to the uncertainty of \dmstruct\ ($\pm0.6\,\mdmunits$). There is a sudden jump to higher time lag values at $\dm\approx564.3\,\mdmunits$, which is reflected in the behaviour of the ACFs in Figure \ref{fig:acf_dms}. The bottom panel shows the frequency-averaged burst profiles and waterfall plots de-dispersed to the relevant DMs. The resolution of the spectra is decimated to 256 channels. Note that the first panel corresponds to the results from \code{DM\_phase}. More sub-components appear to be resolved, as evidenced by the extra peak in the profile.}
\label{fig:acf_and_burst}
\end{figure}

We use the TOP algorithm and verify the results by fitting tangents to the ACFs by eye. In Figure \ref{fig:acf_dms} we compare the ACF of burst 11 for $\Updelta\dm=\pm1\,\mdmunits$ offset from the maximised $\mdmstruct=563.7\,\mdmunits$, where \dmstruct\ is shown to have the smallest time lag. Figure \ref{fig:acf_and_burst} shows the corresponding dynamic spectra and frequency-averaged burst profiles (lower panel), and the DM vs time lag ($\Updelta t_s$) obtained via the TOP algorithm (upper panel).

Final results were obtained by interpolating the DM vs time lag (denoted $f_\text{ACF}$), where \dmstruct\ corresponded to the minimum time lag of the curve (Figure \ref{fig:acf_and_burst}). Akin to the uncertainty estimation technique used in \code{DM\_phase} (see next subsection), the standard deviation was calculated via the Taylor series:
\begin{equation}
\sigma_\text{DM} = \sqrt{\left\lvert \frac{2\sigma_{f\text{\tiny{ACF}}}^2}{f_\text{ACF}''(\mdmstruct)} \right\lvert} \;\;,
\end{equation}
where $\sigma_f$ is given by the residuals of the interpolation.

\subsection{Coherent power spectra}
The \code{DM\_phase} algorithm finds the structure-optimised DM of a burst by maximising the coherent power across the bandwidth \citep{Seymour_2019}:
\begin{equation}
    \label{eq:P_dCo}
    P(\omega,DM) = \omega^2\left\lvert\int\frac{\mathcal{F}\left[ D(t',f) \right]}{ \left\lvert\mathcal{F}\left[ D(t',f) \right]\right\lvert} \mathrm{d}f\right\lvert^2 \;\; ,
\end{equation}
where $\mathcal{F}$ denotes the Fourier transform, $D(t',f)$ is the dynamic spectrum as a function of emission frequency and time, and $\omega$ is the Fourier frequency. Uncertainties are found by converting the standard deviation of the coherent power spectrum into a standard deviation in DM via the Taylor series. For further detail, refer to the {\code{DM\_phase} Github documentation}\footnote{Available at \url{https://github.com/danielemichilli/DM_phase/tree/master/docs}.} and Seymour et al. (\textit{in prep}).

\section{Results and Discussion}
\label{sec:results_discuss}
\begin{figure*}
\subfloat{
	\includegraphics[width=0.7\linewidth]{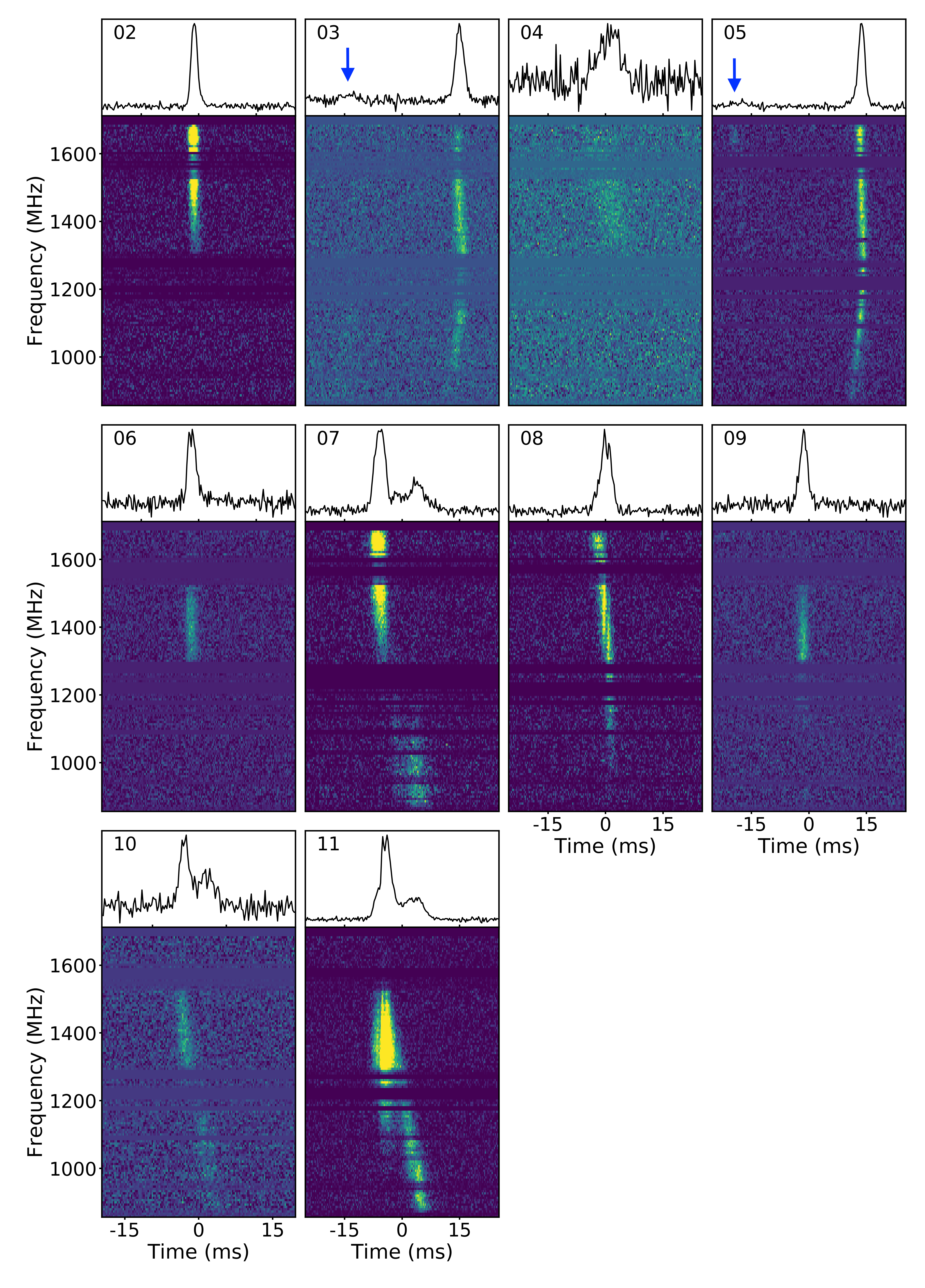}
	}
\caption{Dynamic spectra of the bursts detected with MeerKAT on the 10th of September 2019. The top panels show the frequency-averaged burst profile. The bottom panels show the frequency spectra with the resolution of each burst decimated to 256 channels to enhance visibility. The time resolution of the bursts is $306.24\,\upmu$s. The RFI was removed from the data manually. The flux density scale is uncalibrated and shown in arbitrary units. Please refer to \citet{2020MNRAS.496.4565C} for full details. The bursts are de-dispersed to the structure-optimised DMs given in Table \ref{tab:dm_struct} (the mean of ACF and \code{DM\_phase}). Burst 01 is not shown, as a structure-optimised DM could not be established. Burst 02 is de-dispersed to $564.6\,\mdmunits$, burst 03 to $565.9\,\mdmunits$, burst 04 to $572.4\,\mdmunits$, burst 05 to $564.5\,\mdmunits$, burst 06 to $563.1\,\mdmunits$, burst 07 to $563.0\,\mdmunits$, burst 08 to $563.6\,\mdmunits$, burst 09 to $565.1\,\mdmunits$, burst 10 to $563.4\,\mdmunits$, and burst 11 to $563.3\,\mdmunits$.}  
\label{fig:gallery}
\end{figure*}

The structure-optimised DMs are presented in Table \ref{tab:dm_struct}. The frequency spectra (`waterfall' plots) are shown in Figure \ref{fig:gallery}. Where the structure-optimised DMs given by the ACF and \code{DM\_phase} methods agree, bursts are de-dispersed to the mean of the two results. Where they differ, the most likely candidate DM is used (as discussed in Section \ref{ssec:compare}). Section \ref{ssec:compare} provides a comparison of the two techniques, after which the average DM for the epoch is calculated. A number of caveats in determining \dmstruct\ are highlighted here. The burst properties are then presented and possible implications are discussed.

\begin{table}
	\centering
	\caption{Structure-optimised DMs for the 11 FRB\,121102 bursts. Due to their low fluxes, including/excluding the precursors in the analysis for bursts 03 and 05 did not affect the value of \dmstruct. For bursts 07, 08 and 10, \code{DM\_phase} gave multiple possible values, as discussed in Section \ref{ssec:compare}. The last column gives the best estimate for each burst. Where the ACF method and \code{DM\_phase} results agree, the mean of the results is used, rounded up to the nearest decimal value. Where results disagree, the most likely value is chosen, as discussed in Section \ref{ssec:compare}. Asterisks denote the selected sample of best estimates used in the second calculation of the average DM.}
	\label{tab:dm_struct}
	\begin{tabular}{ l c c c}
        \hline\hline
        Burst & ACF Method & DM\_phase & Value chosen \\
        & (\dmunits) & (\dmunits) & (\dmunits) \\
        \hline
        01 & --- & --- \\ 
        02 & $564.1\pm0.3$ & $565.1\pm0.4$ & $564.6\pm0.4^{*}$ \\
        03 & $566.0\pm0.2$ & $565.8\pm0.2$ & $565.9\pm0.2$  \\
        04 & $572.0\pm0.8$ & $572.7\pm0.3$ & $572.4\pm0.6$  \\
        05 & $564.5\pm0.3$ & $564.4\pm0.3$ & $564.5\pm0.3$  \\
        06 & $562.8\pm0.9$ & $563.4\pm0.7$ & $563.1\pm0.8$  \\
        07 & $563.1\pm0.4$ & $562.9\pm0.2^{*}$ & $563.0\pm0.3$ \\
          &                & $564.4\pm0.2$  & \\
          &                & $565.6\pm0.2$  & \\
        08 & $564.4\pm0.4$ & $563.6\pm0.1^{*}$ & $563.6\pm0.1$ \\
          &                & $564.9\pm0.6$  & \\
        09 & $565.0\pm0.9$ & $565.2\pm1.1$  & $565.1\pm1.0$ \\
        10 & $563.3\pm0.4$ & $563.6\pm0.4$ & $563.5\pm0.4$ \\
           &               & $565.8\pm0.4$ &  \\
        11 & $563.7\pm0.6$ & $562.8\pm0.3^{*}$ & $563.3\pm0.5$ \\ 
        \hline
    \end{tabular}
\end{table}

\subsection{Comparison of techniques}
\label{ssec:compare}
\begin{figure}
\subfloat{
	\includegraphics[width=1\linewidth]{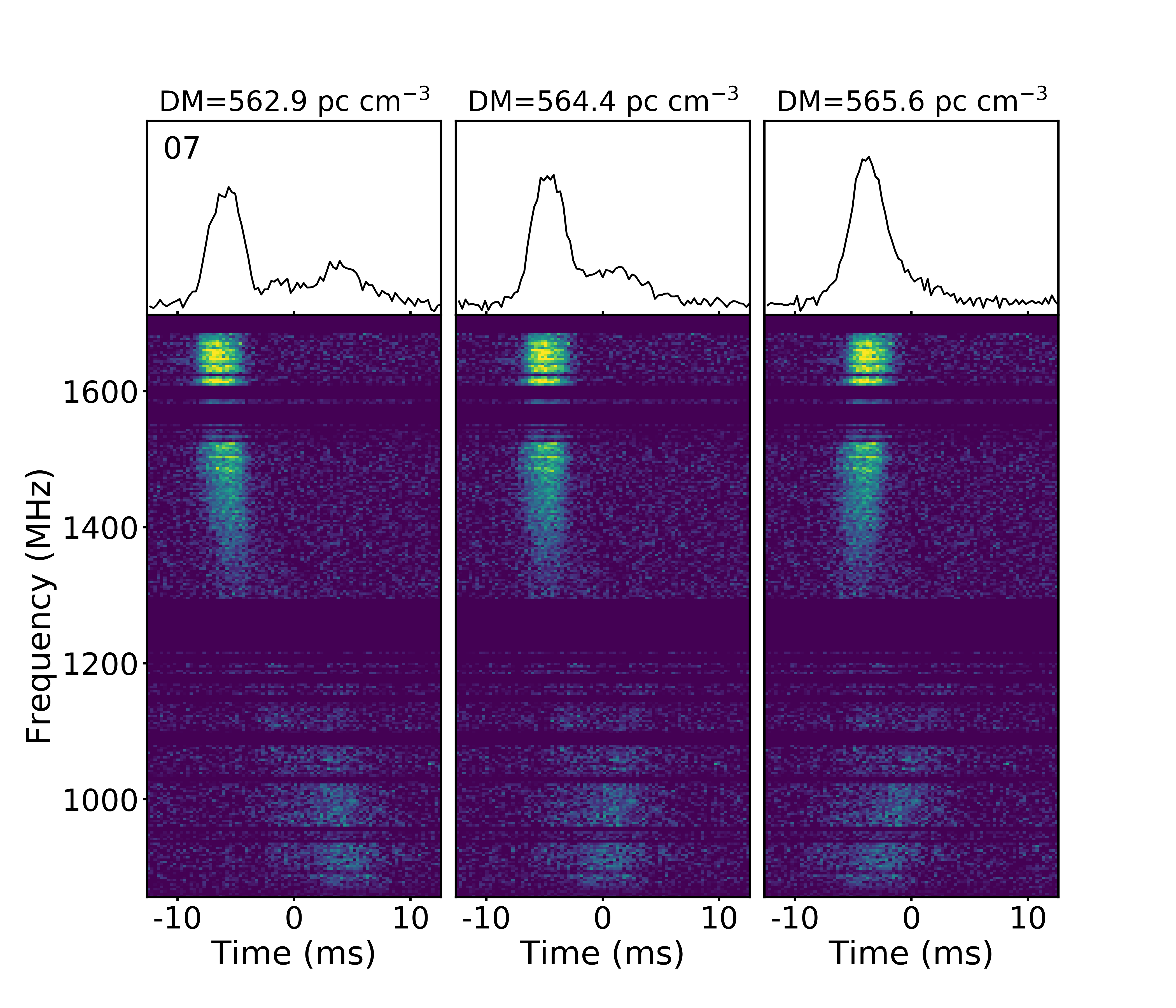}
	}
\caption{Burst 07 at the structure-optimised DMs identified by \code{DM\_phase}. The resolution of the spectra is decimated to 256 channels. The first panel ($\mdmstruct=562.9\,\mdmunits$) agrees with the ACF method ($\mdmstruct=563.1\pm0.4\,\mdmunits$). Three (possibly four) sub-bursts are evidenced, which march down in frequency. The profile structure then decreases as the lower frequency bursts begin to sweep under the main burst.
}
\label{fig:07_wfall}
\end{figure}

\begin{figure}
\subfloat{
	\includegraphics[width=1\linewidth]{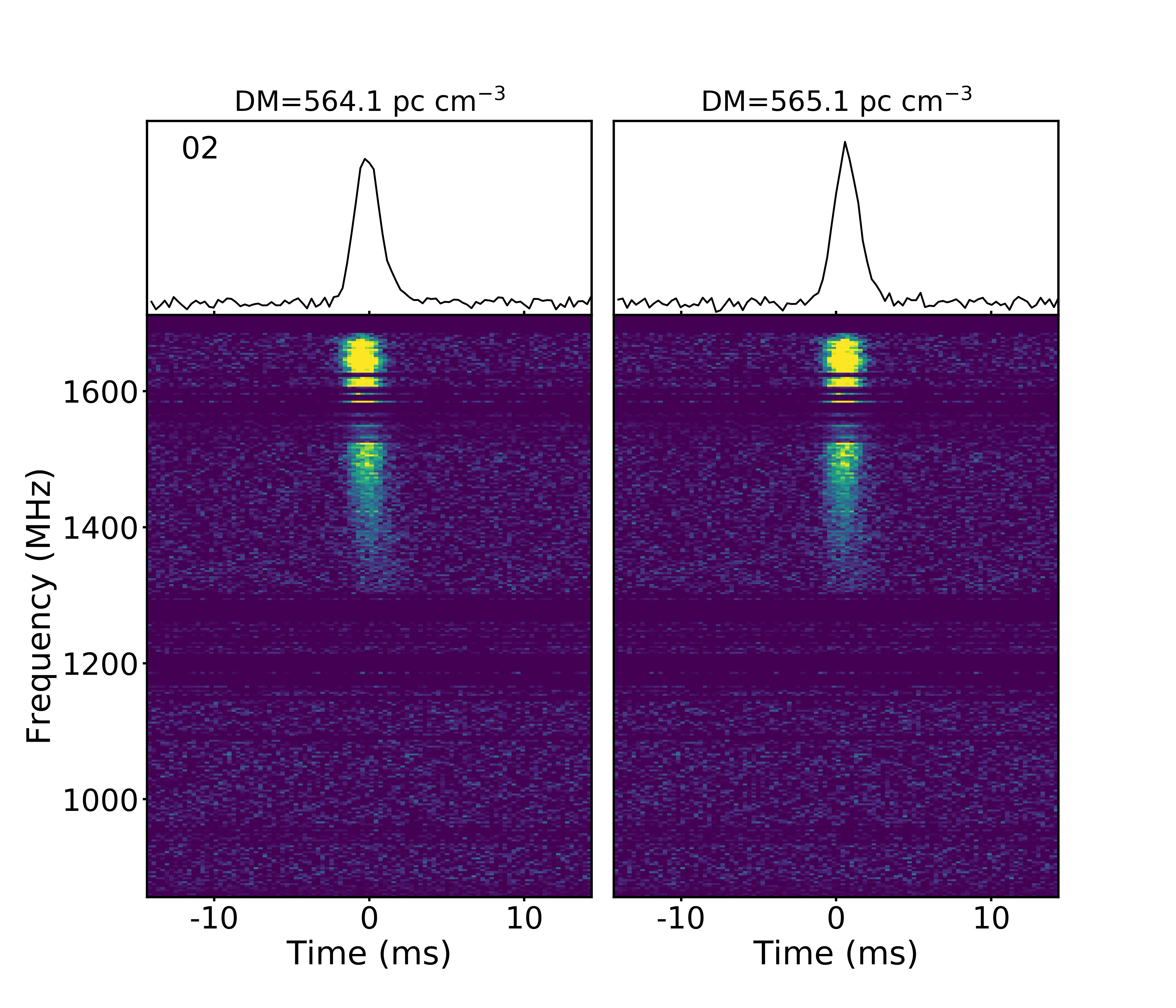}
	}
\caption{Burst 02 at the structure-optimised DMs identified by the ACF method (first pannel) and \code{DM\_phase} (second pannel). There is little observable change in the structure. The resolution of the spectra is decimated to 256 channels.
}
\label{fig:02_wfall}
\end{figure}

The results from the ACF method and \code{DM\_phase} largely agree to within a 1$\sigma$ confidence level, and both methods did not find structure when bursts were particularly faint (i.e. burst 01 and the precursor of burst 03, the latter of which is discussed in Section \ref{ssec:burstpair}). \code{DM\_phase} gives multiple values for bursts 07, 08 and 10, which necessitates further investigation. The multiple values are evidenced by multiple peaks in the coherent power spectra. For burst 07, three values were determined for \dmstruct\ using \code{DM\_phase} (Figure \ref{fig:07_wfall}). The first, with $\dm=562.9\pm0.2\,\mdmunits$, agrees with the ACF method. At this DM value, there are at least three distinct sub-bursts. We take this to be the structure-optimised DM. For the next two \code{DM\_phase} values for \dmstruct, the sub-bursts begin to align with the main burst in time, and the structure in the frequency-averaged profile diminishes. While the results for burst 11 agree within the uncertainty margins, the burst profiles look significantly different at the central values (Figure \ref{fig:acf_and_burst}). We argue that the most representative structure-optimised DM is given by \code{DM\_phase} ($\mdmstruct=562.8\pm0.3\,\mdmunits$), where one can see an additional peak in the profile. Further, at this DM a bright sub-component of the second burst aligns with the main burst. The results from ACF and \code{DM\_phase} also differ for burst 02. Here it is unclear which is most likely, as the burst profile changes so little (Figure \ref{fig:02_wfall}). We take the structure-optimised DM to be the mean of the two methods.

\begin{figure}
\subfloat{
	\includegraphics[width=1\linewidth]{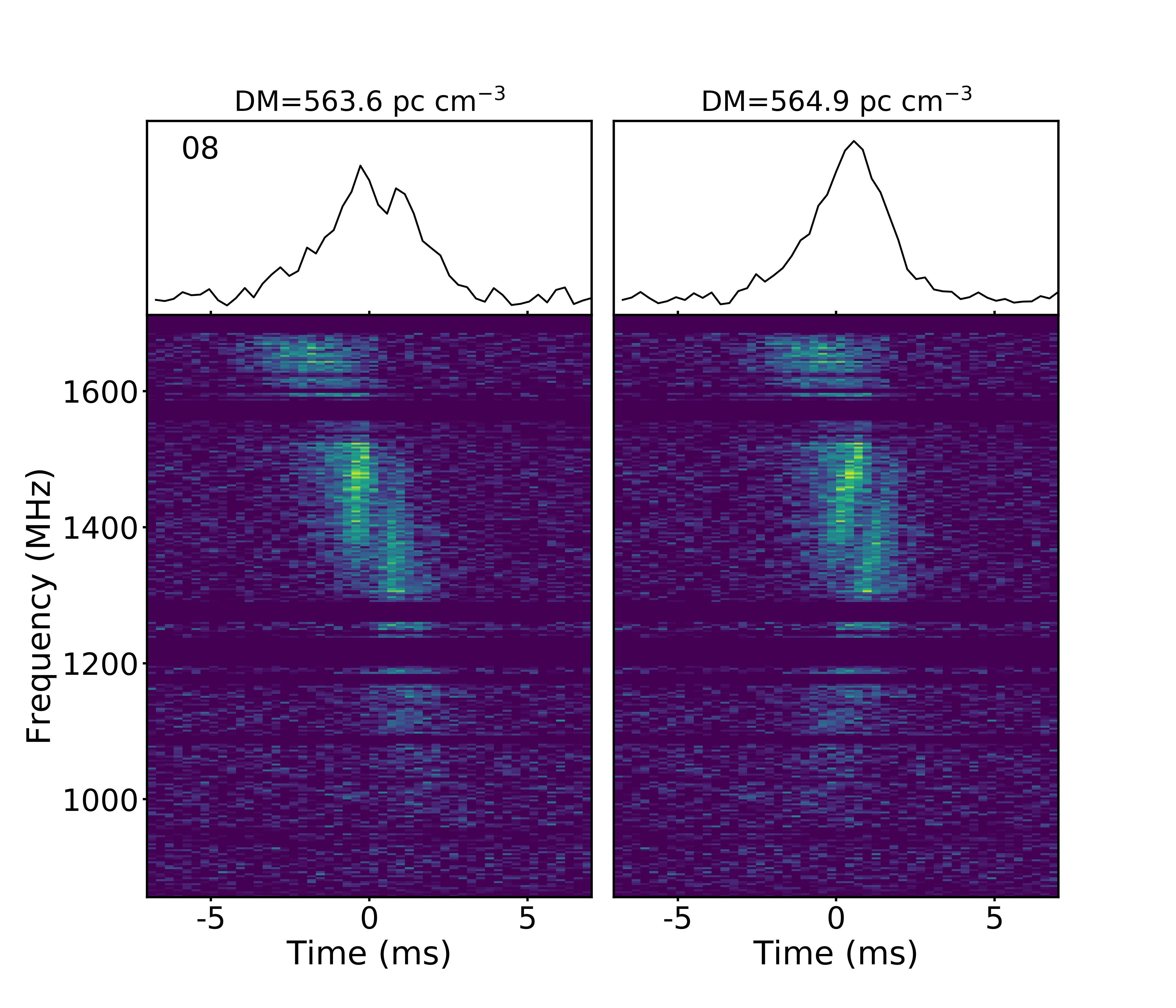}
	}
\caption{Burst 08 at the structure-optimised DMs identified by \code{DM\_phase}. The resolution of the spectra is decimated to 256 channels. The second panel ($\mdmstruct=564.9\,\mdmunits$) agrees with the ACF result ($\mdmstruct=564.6\pm0.4\,\mdmunits$), however here the two sub-bursts are not reflected in the frequency-averaged profile and the bursts overlap each other. At $\mdmstruct=563.6\,\mdmunits$ the sub-bursts are distinct in the profile and show a downward frequency drift in the waterfall plot.}
\label{fig:08_wfall}
\end{figure}

\begin{figure}
\subfloat{
	\includegraphics[width=1\linewidth]{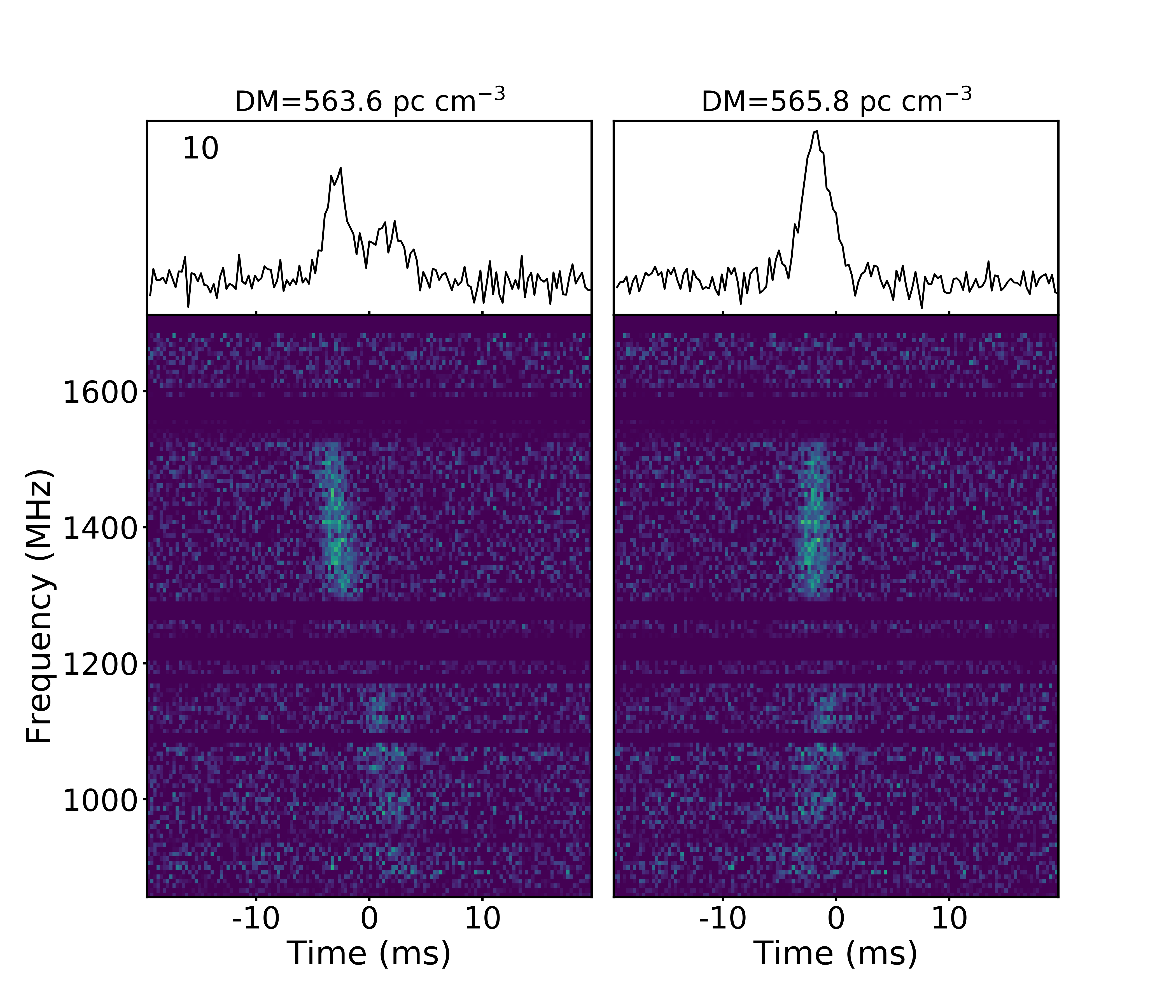}
	}
\caption{Burst 10 at the structure-optimised DMs identified by \code{DM\_phase}. The resolution of the spectra is decimated to 128 channels to enhance visibility. The first panel ($\mdmstruct=563.6\,\mdmunits$) agrees with the ACF result ($\mdmstruct=563.3\pm0.4\,\mdmunits$). In this case, there are two sub-bursts. In the second panel, the lower burst sweeps under the upper burst. The behaviour and appearance of burst 10 is similar to that of burst 07. Arguably, the missing frequency bands at ${\sim}1250$\,MHz may create the illusion of two sub-bursts.}
\label{fig:10_wfall}
\end{figure}

Figure \ref{fig:08_wfall} shows the two \code{DM\_phase} values for \dmstruct\ for burst 08. The ACF method agrees with the second result, where $\mdmstruct=564.9\pm0.6\,\mdmunits$. Here, however, the profile structure has been washed out. At $\mdmstruct=563.6\,\mdmunits$, one can see the two peaks from the sub-bursts. Burst 10 is shown in Figure \ref{fig:10_wfall}. Here, it is unclear whether or not the burst consists of two sub-bursts -- the missing frequency bands at ${\sim}1250$\,MHz could indicate the appearance of two distinct bursts. Looking at the second panel, the top half of the burst does not align with the bottom half, which may suggest that they are sub-bursts. The behaviour and appearance of burst 10 is also similar to that of burst 07. This may imply $\mdmstruct=563.6\pm0.4\,\mdmunits$. One may also argue for the lower DM value by noting that it is better in line with previous DM measurements of FRB\,121102 \citep{Hessels2019,Josephy_2019,Oostrum_2020}. Ultimately, however, the result is ambiguous.

In summary, as a result of the comparison, we note that (i) \code{DM\_phase} occasionally gives multiple possible values for \dmstruct, and it is important to manually check these, as the highest peak in the power-DM function does not necessarily correspond to the structure-optimised DM; (ii) the ACF method failed to identify structure where the burst separation is less than a millisecond (e.g. burst 08) and the analysis requires more manual intervention.

While the two definitions of `maximum structure' give results that are largely consistent with each other, ambiguity still exists within the metric. For example, at 2$\sigma$, the two solutions for burst 8 are compatible, but visual inspection shows they are clearly alternative to each other. Care should be taken when performing these types of analyses and results should be accepted with a measure of caution. Going forward, it will be interesting to compare these methodologies with those of \citet{2018ApJ...863....2G} and \citet{Hessels2019} for bursts with higher time resolution data.

\subsection{Average DM variation}
The average structure-optimised DM of the epoch is calculated in two ways. In the first, the average is taken over the 10 bursts, weighted by the errors, to give $\mdmstruct=564.8\pm0.6 (\text{sys}) \pm 2.5 (\text{stat}) \,\mdmunits$ and $\mdmstruct=564.4\pm0.6 (\text{sys}) \pm 2.9 (\text{stat}) \,\mdmunits$ using the ACF method and \code{DM\_phase}, respectively. The first uncertainty is the systematic uncertainty given by the respective methods and the second is the statistical uncertainty given by the standard deviation of the data. The values of \dmstruct\ for some of the bursts fall outside of this region; most notable of which is burst 04, which is ${\sim}8\,\mdmunits$ greater than the average. We attribute this difference to insufficient S/N. Unresolved components in the bursts may also significantly influence the resultant \dmstruct.
We thus urge caution when interpreting the DM change between bursts in this, and for that matter any, sample. We also note that the errors given by both methods are under-representations of the true uncertainty on the measurement, as they do not take into account potentially unresolved components. As such, even individual results with small uncertainties should be closely examined. Good examples of this are bursts 03 and 05, whose ambiguity is discussed in Section \ref{sssec:dmmainbursts}.

\begin{figure*}
\subfloat{
	\includegraphics[width=0.7\linewidth]{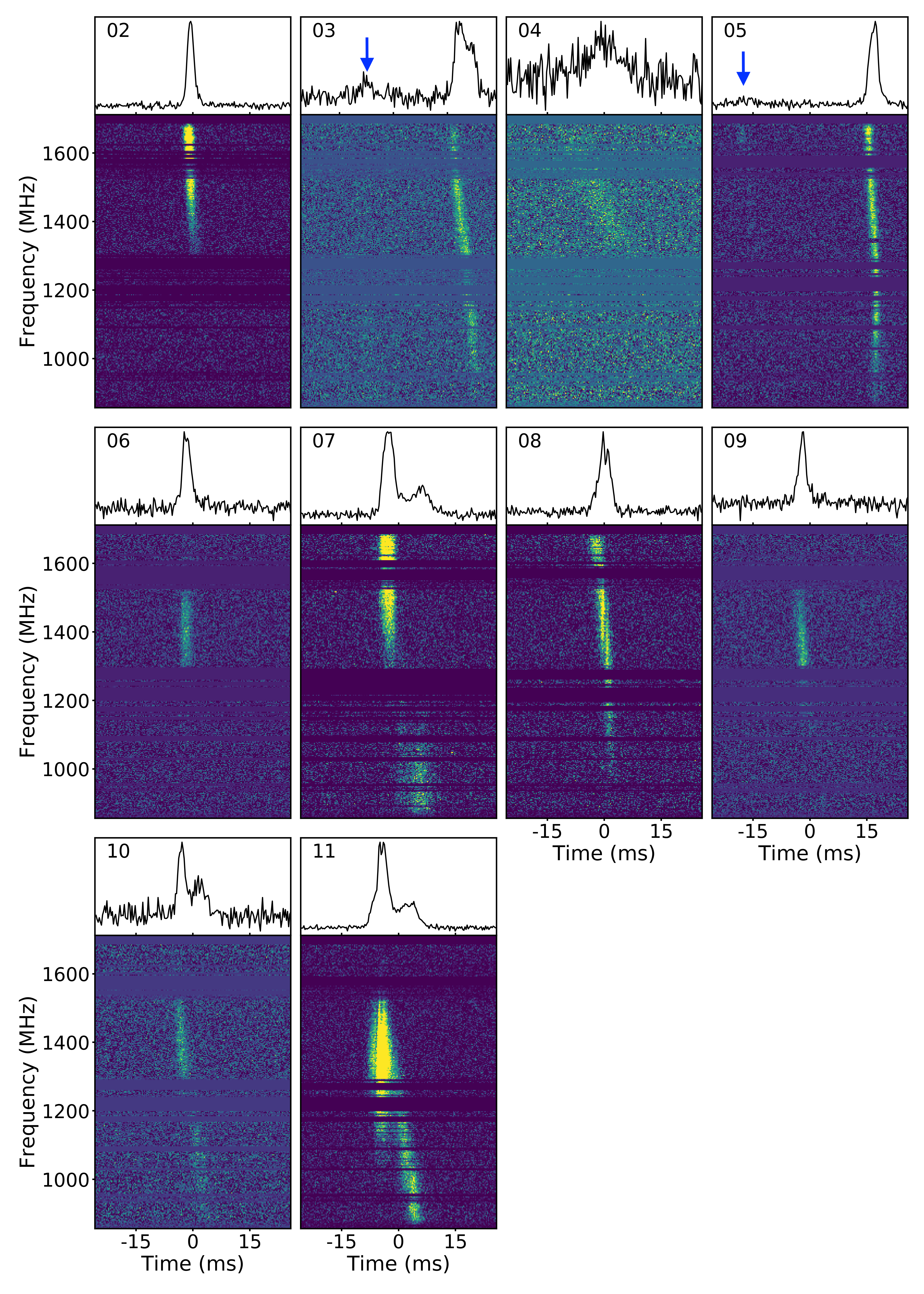}
	}
\caption{Bursts de-dispersed to an average structure optimised DM of ${\sim}563.5 \,\mdmunits$. Note the difference in the behaviour of the main bursts of 03 and 05 from that shown in Figure \ref{fig:gallery}. Instead of showing an apparent deviation from the $t\sim\nu^{-2}$ relationship, the middle section of the main bursts are misaligned, and are thus possibly made up of unresolved downward drifting sub-bursts.}
\label{fig:ave_gall}
\end{figure*}

Establishing a reliable mean DM for the epoch may best be achieved by only considering bursts whose sub-components appear to be reasonably resolved. As such, we recalculate the average DM with a selected sample of bursts and their best estimates.  The final data set consists of burst 02 (mean of ACF and \code{DM\_phase}), burst 08 (\code{DM\_phase}), burst 07 (\code{DM\_phase}) and burst 11 (\code{DM\_phase}). This gives a structure-optimised DM of $563.5\pm 0.2 (\text{sys}) \pm 0.8 (\text{stat})\,\mdmunits$. Figure \ref{fig:ave_gall} shows the bursts de-dispersed to $563.5\,\mdmunits$. An important question then is whether this single DM creates a cohesive picture of the burst sample. This is discussed in Section \ref{sssec:dmmainbursts}.

\begin{figure}
\subfloat{
	\includegraphics[width=\linewidth]{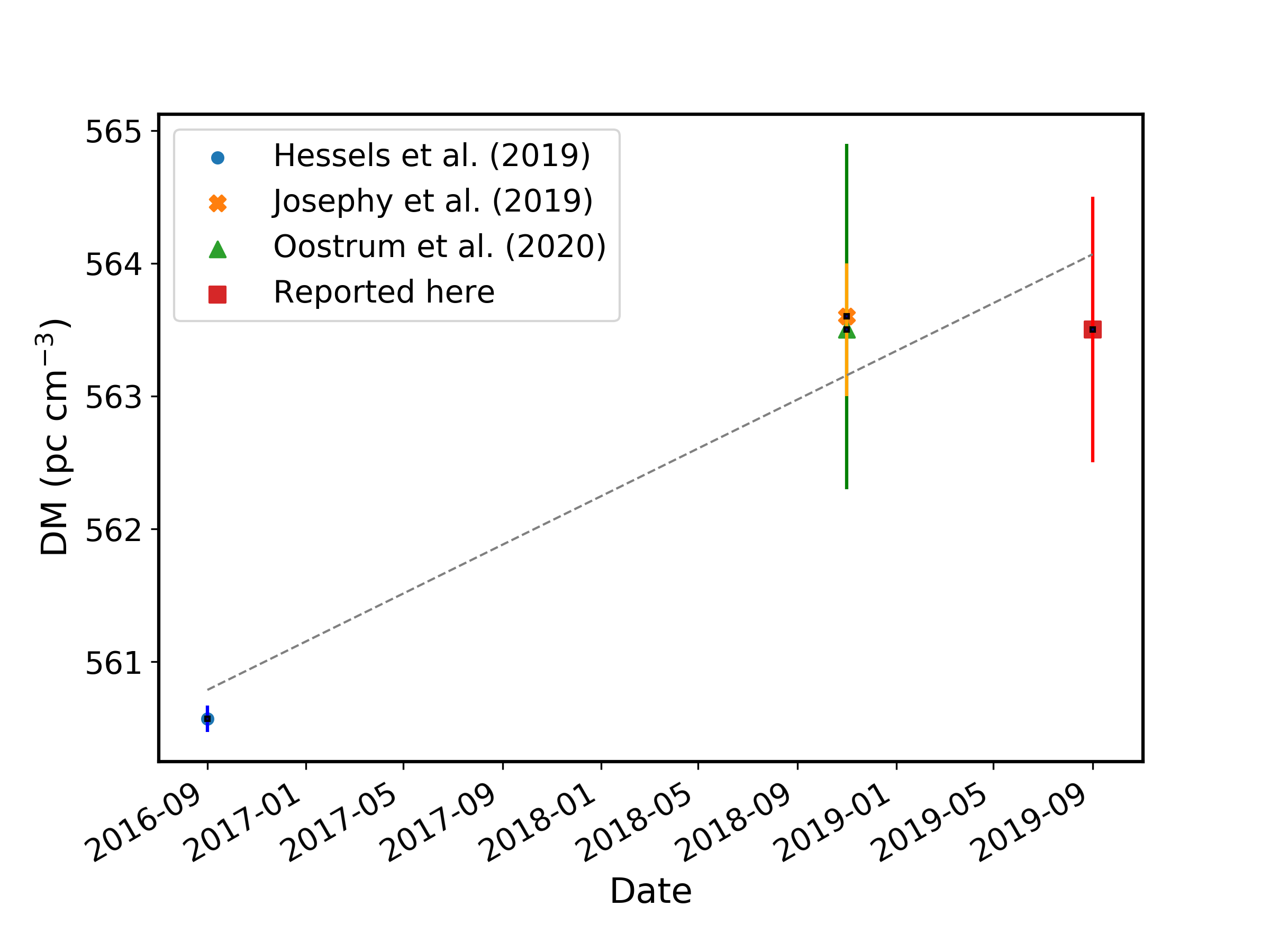}
	}
\caption{Structure-optimised DMs measured for FRB\,121102 between 2016 and 2019. The dashed grey line shows the linear interpolation, which gives an average increase of ${\sim}\,1\, \mdmunits$ per year. \citet{Hessels2019} and \citet{Josephy_2019} use the maximum steepness method to determine \dmstruct, and \citet{Oostrum_2020} use \code{DM\_phase}.}
\label{fig:dm_change}
\end{figure}

The average structure-optimised DM is consistent with 2018 observations ($563.6\pm0.5 \,\mdmunits$; \citealt{Josephy_2019} and $563.5\pm1.3 \,\mdmunits$; \citealt{Oostrum_2020}) taken 1 year prior. The uncertainties make it unclear whether the average DM has indeed remained constant over this period, or whether it has increased or even decreased. A linear interpolation with 2016 observations \citep[${\sim}560.6 \,\mdmunits$;][]{Hessels2019} reveals an average increase of ${\sim}\,1\,\mdmunits$ (Figure \ref{fig:dm_change}). This is roughly consistent the ${\sim}1{-}3 \,\mdmunits$ increase from 2012 to 2016 reported by \citet{Hessels2019}, however more data is needed in our case to confirm whether the increase is indeed secular. There are a number of scenarios that may account for the apparent trend. A persistent increase in DM may, for example, be attributed to a young neutron star whose supernova ejecta expands into a high density interstellar medium \citep[ISM; ][]{Yang_2017,Piro_2018}. The FRB may also be associated with a young star whose ionisation drives outward expansion into a surrounding H\,{\small II} region \citep{Yang_2017}. Alternatively, if the FRB source is moving rapidly through an H\,{\small II} region due to -- for instance -- a supernova kick, the DM may increase or decrease depending on the direction of the kick \citep{Yang_2017}. In the magnetar flare model by \citet{margalit+18}, the increase in DM may be attributed to the photoionisation of neutral gas by the UV and X-ray radiation from the shock. Here, an increase of $0.01{-}1\,\mdmunits$ is expected on a time scale of days to months. For an in-depth discussion on the DM and RM evolution of FRB\,121102 in the context of the supernova remnant models by \citet{Piro_2018} and \citet{margalit+18}, see \citet{hilmarsson+20}.

Should the DM be shown to decrease in the future, plasma lensing may also be accountable (although the scenarios mentioned above would not be ruled out by this). In this case, plasma lensing would be local to the source (e.g. a nebula) or the host galaxy \citep[e.g. AGNs;][]{Cordes_2017}. It has been shown that non-local propagation effects, such as from Hubble expansion, gas density fluctuations in large-scale structure and gravitational potential fluctuations, cannot account for the observed DM variations of FRB\,121102 \citep{Yang_2017}.

\subsection{Sub-bursts}
Bursts 07 and 11 (and possibly 10) have a bifurcating structure around $1250$\,MHz.  Similar behaviour (at a different central frequency) has been observed in FRB\,121102 before \citep[burst GB-01;][]{Hessels2019} and in FRB\,180916.J0158+65 \citep[burst 11;][]{2020arXiv200714404M}, where the right-most component of each burst appears to follow a different DM to the previous components. Particularly notable is latter: burst 11 in \citet{2020arXiv200714404M}, where a bright component that aligns with the previous sub-bursts is embedded in the right-most sub-burst. This presents the possibility that the sub-bursts of bursts 07 and 11 are not single sub-bursts with a different DM, but rather comprise multiple unresolved components that drift down in frequency.

We investigate the apparent change in DM between the sub-bursts by splitting the spectra at 1250\,MHz and 1100\,MHz, respectively. We note that the RFI affected (and thus removed) frequency bands at ${\sim}1250$\,MHz and the overlapping frequencies of the different components may affect the DM results. For burst 07, the DM for the higher frequency sub-burst is ${\sim}1\,\mdmunits$ lower than the lower frequency sub-burst (a 1$\sigma$ difference; $563.3\pm0.7$\,\dmunits\ vs $564.4\pm0.4$\,\dmunits, using \code{DM\_phase}). For burst 11, the higher frequency sub-burst is ${\sim}2\,\mdmunits$ lower than the lower frequency sub-burst ($562.7\pm0.4$\,\dmunits\ vs $564.9\pm0.5$\,\dmunits, using \code{DM\_phase}). Interestingly, we note that if the bursts are multiple images caused by plasma lensing, the predicted observed difference in DM values could be as large as ${\sim}1\,\mdmunits$ \citep[][also see Section \ref{sssec:plasmalens}]{Cordes_2017}. 

Other mechanisms can in principle be invoked to explain a difference in DM: e.g. the sub-bursts could be emitted from different parts of the magnetosphere, or they could be observed along different sight-lines through dense plasma \citep{2020ApJ...891L..38C,2020MNRAS.497.3335D} -- for example the line-of-sight through a nebula will vary depending on the neutron star rotation phase at the time of emission \citep{2020ApJ...899L..21S}. However, the differential DMs expected in these cases are too small to account for those observed in bursts 07 and 11. It will be interesting to see in higher resolution data going forward whether similar sub-bursts truly do misalign with previous sub-bursts or if the effect is a result of unresolved downward drifting sub-structure. This clearly has implications on the DM of FRB\,121102 for this epoch and other epochs. If the right-most sub-bursts are made up of unresolved sub-bursts that align with the higher frequency sub-burst, then the DM of the burst is best described by the DM of the higher frequency sub-burst.

\begin{figure}
\subfloat{
	\includegraphics[width=0.54\linewidth, trim=0.0cm 2.5cm 0cm 2.0cm, clip=True]{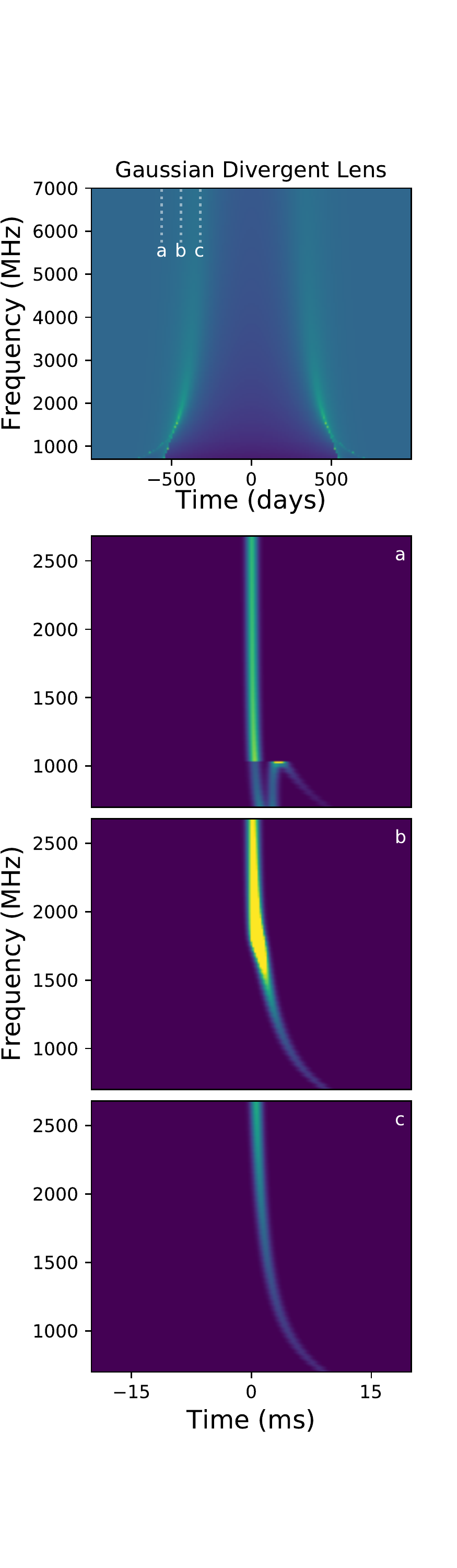}
	\includegraphics[width=0.442\linewidth, trim=1.6cm 2.5cm 0cm 2.0cm, clip=true]{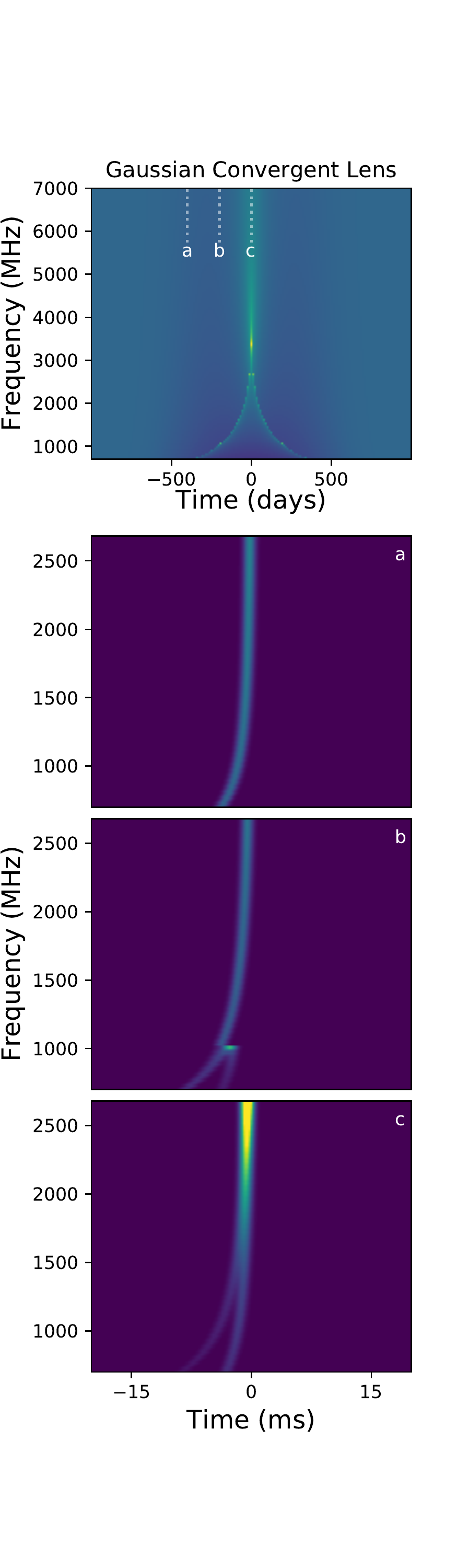}
	}
\caption{Lensing simulations using an overdense (\textit{left}) or underdense (\textit{right}) Gaussian lens, as described in Section \ref{sssec:plasmalens}.  Top panels show the magnification $\mu(t, f)$, with a logarithmic colour bar extending from $10^{-1} - 10^{2}$. The bottom panels show a mock FRB (modelled as an achromatic Gaussian ($\mu(f)=1$ with 0.5\,ms width) with magnification, geometric time delay, and dispersive delays of the lensing field at a given time applied. The colourbar is saturated to magnifications between 0 and 5.}
\label{fig:lens_sim}
\end{figure}

\subsection{Burst pairs}
\label{ssec:burstpair}

In bursts 03 and 05, a bright (main) burst is preceded by a faint (precursor) burst, with a separation time of ${\sim}28$\,ms and ${\sim}34$\,ms, respectively. FRB burst pairs are not connected by an emission bridge. This distinguishes burst pairs from sub-bursts \citep[e.g.][]{2020MNRAS.497.3335D}. 

Numerous burst pairs have been observed in FRB\,121102 before, as well as in other FRBs (both repeating and apparently non-repeating). A summary is provided in Table \ref{tab:pairs}.
\begin{table}
	\centering
	\caption{Burst pairs observed in FRBs. Where bursts are from repeaters, the burst name for the individual burst (given in italics) follows the naming convention of the relevant paper. If there is no convention, the name corresponds to the observation number $X$ in the relevant paper as B$X$. Apparently one-off FRBs are named as per usual.}
	\label{tab:pairs}
	\begin{tabular}{ c c c c}
        \hline\hline
        FRB & Separation & Reference \\
        & (ms) \\
        \hline
        FRB\,121102: &  &  \\
        \textit{GB 1/2} & ${\sim}37$ & \citet{Scholz_2017} \\
        \textit{B10/11} & ${\sim}34$ & \citet{Hardy_2017} \\
        \textit{B35/36} & ${\sim}26$ & \citet{2019ApJ...877L..19G} \\
        \textit{B20/21} & ${\sim}38$ & \citet{2020arXiv200803461C} \\
        \textit{Burst 2} & ${\sim}17$ & \citet{2020arXiv200903795R} \\
        \textit{Burst 03} & ${\sim}28$ & \citet{2020MNRAS.496.4565C}; here \\
        \textit{Burst 05} & ${\sim}34$ & \citet{2020MNRAS.496.4565C}; here \\
        \hline
        FRB\,180916.J0158+65: &  &  \\
        \textit{181019} & ${\sim}60$ & \citeauthor{andersen_2019} \\
        & & \citeyear{andersen_2019}\\
        \textit{191219A/B} & ${\sim}60$ & \citet{Chawla_2020}; \\
        &  & \citeauthor{Amiri_2020} \\
        &  & \citeyear{Amiri_2020} \\
        \textit{200620} & ${\sim}90$ & \citet{2020arXiv200714404M} \\
        \hline
        FRB\,190212.J18+81: &  &  \\
        \textit{190213} & ${\sim}19$ & \citet{Fonseca_2020} \\
        \hline
        FRB\,200428: &  &  \\
        \textit{B1/2} & ${\sim}29$ & \citet{Bochenek_2020}; \\
        &  &  \citeauthor{Andersen_2020} \\
        &  &  \citeyear{Andersen_2020} \\
        \hline
        FRB\,181112 & ${\sim}0.5$ & \citet{2019Sci...366..231P}; \\
        &  &  \citet{2020ApJ...891L..38C} \\
        \hline
        FRB\,190102 & ${\sim}0.1$ & \citet{2020MNRAS.497.3335D} \\
        \hline
        FRB\,190611 & ${\sim}0.7$ & \citet{2020MNRAS.497.3335D} \\
        \hline
    \end{tabular}
\end{table}

We note that the waiting time between the burst pairs of apparently one-off bursts are significantly shorter than those of repeaters. Due to the small sample size, this may just be coincidence.

\subsubsection{Burst Envelope}
Whether or not burst pairs are independent events (or even echoes; Section \ref{ssec:pol}) is currently an open question. \citet{2020arXiv200803461C}, for example, propose that burst pairs may be broad bursts with only two resolvable components. In the case of a neutron star, if the source is active as a radio emitter for a duration similar to its rotation period, then we may expect to see pre- and postcursor bursts, as well as, occasionally, both of them. To date, no such triplets have been observed.

The longest duration of a single burst reported for FRB\,121102 is $39\pm2$\,ms \citep[burst B31;][]{2020arXiv200803461C}, which is comparable to the total time scales of bursts 03 and 05 (${\sim}37$\,ms and ${\sim}39$\,ms, respectively). As such, it is feasible that the bursts occurred within the same burst envelope, and hence that burst 03 shows upward drift, i.e. the main burst of 03 arrives at a higher frequency than the precursor (Figure \ref{fig:gallery}). This comparison, however, provides only tenuous evidence.

\subsubsection{The DM of the Main Bursts}
\label{sssec:dmmainbursts}
\begin{figure}
\subfloat{
	\includegraphics[width=1\linewidth]{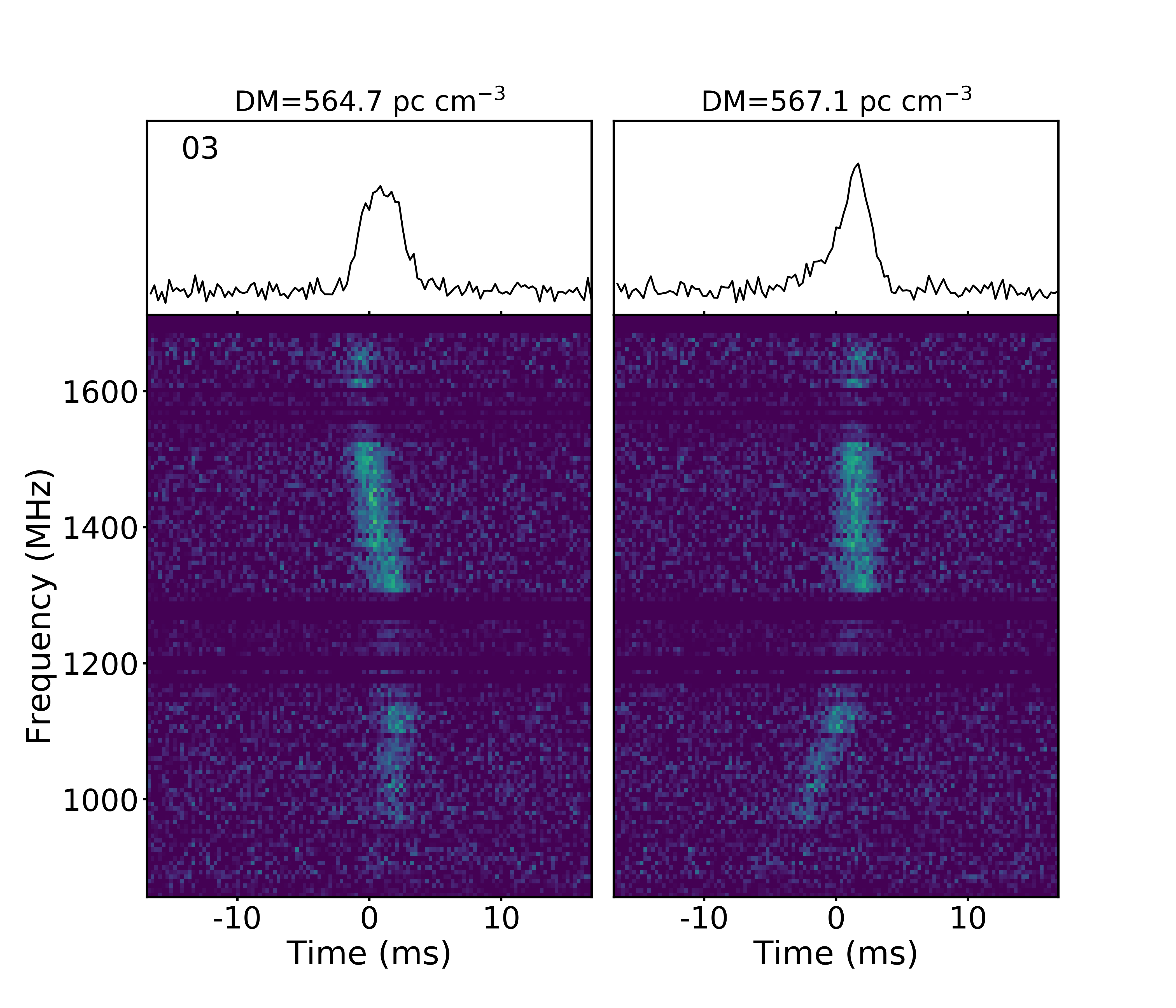}
	}
\caption{Burst 03 de-dispersed to the structure-optimised DMs given by frequency bands below (panel 1) and above (panel 2) 1210\,MHz. The resolution of the spectra is decimated to 256 channels.}
\label{fig:03_wfall}
\end{figure}

\begin{figure}
\subfloat{
	\includegraphics[width=1\linewidth]{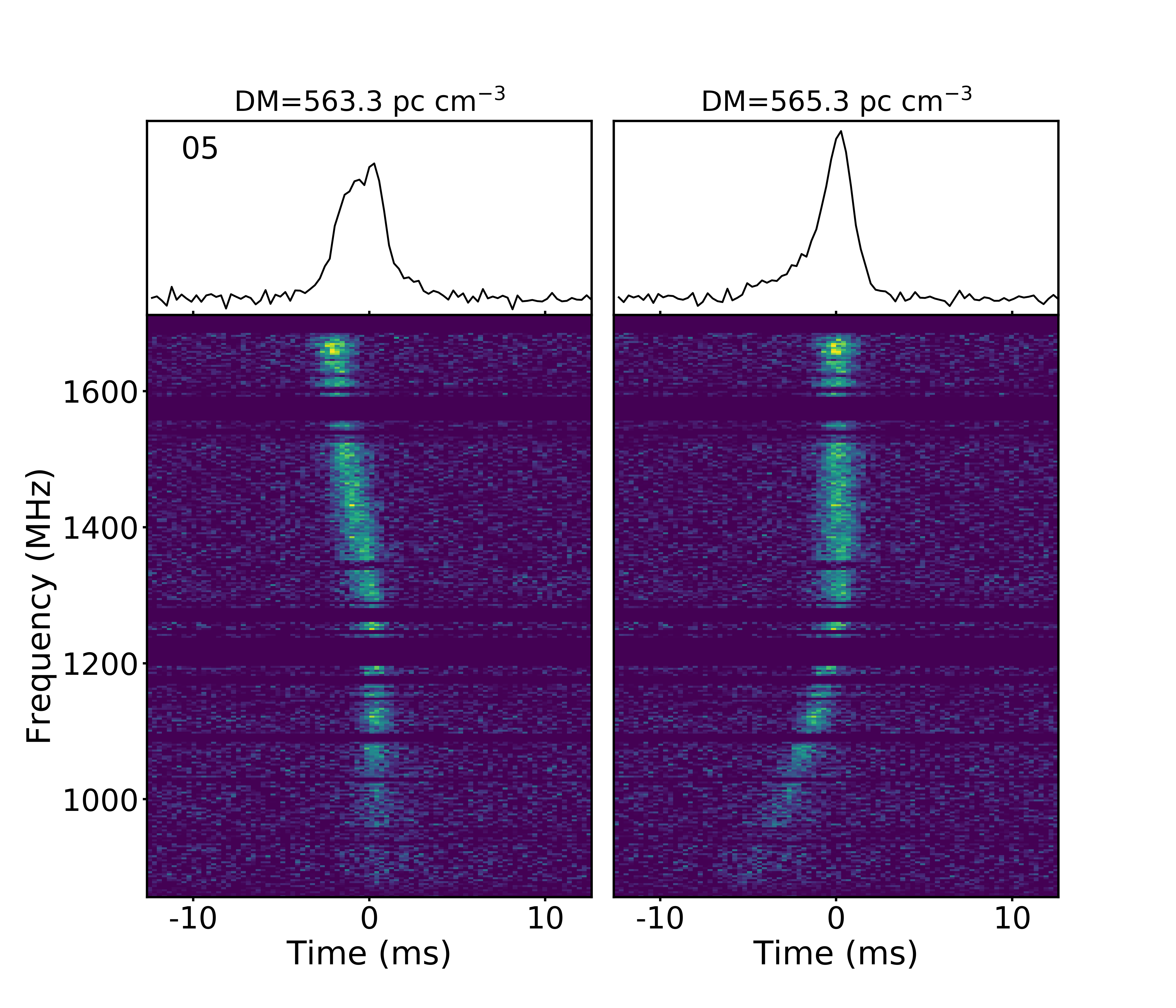}
	}
\caption{Burst 05 de-dispersed to the structure-optimised DMs given by frequency bands below (panel 1) and above (panel 2) 1210\,MHz. The resolution of the spectra is decimated to 256 channels.}
\label{fig:05_wfall}
\end{figure}

When de-dispersed to their structure-maximised DMs, the main bursts of bursts 03 and 05 appear to change behaviour at ${\sim}1250$\,MHz, where the tails abruptly tilt to earlier times (see Figures \ref{fig:03_wfall} and \ref{fig:05_wfall}). In our sample, this feature is exclusive to the bursts with precursors. In particular (using \code{DM\_phase}), the lower and upper frequency bands of burst 03 have $\mdmstruct=564.7\pm0.7\,\mdmunits$ and $\mdmstruct=567.1\pm0.5\,\mdmunits$, respectively; and the lower and upper frequency bands of burst 05 have $\mdmstruct=563.3\pm0.2\,\mdmunits$ and $\mdmstruct=565.3\pm0.2\,\mdmunits$, respectively. This may indicate a deviation from the $\nu^{-2}$ law. On the other hand, it is possible that correctly de-dispersing the lower frequency parts of the bursts may give the most representative DM, even though that component is not dominant over the observed bandwidth. In this case, the upper part of the bursts would comprise unresolved downward drifting sub-bursts (the first panels of Figures \ref{fig:03_wfall} and \ref{fig:05_wfall}). In support of this scenario, the lower DM values are more in line with other bursts in the sample and with the previously reported DM values for FRB\,121102 \citep{Hessels2019,Josephy_2019,Oostrum_2020}.

Interestingly, there is a differential DM of ${\sim}1$\,\dmunits\ between the main bursts of burst 03 and 05, as illustrated by their shapes at the DMs depicted in Figures \ref{fig:03_wfall} and \ref{fig:05_wfall}: burst 03 looks the same as burst 05 when it is de-dispersed to values ${\sim}1$\,\dmunits\ higher than burst 05. This further highlights the challenges in determining an average or representative DM for an epoch -- there may be no single DM that best describes all bursts in a sample, and it is difficult to isolate genuine changes in DMs between bursts.

\subsubsection{Plasma Lensing}
\label{sssec:plasmalens}
Here we consider the potential change in behaviour observed in the main bursts of 03 and 05 when de-dispersed to their individual structure-optimised DMs. The deviation from a $\nu^{-2}$ law could be caused by multi-path propagation, either through geometric delays (caused by the differing path lengths of light across frequency), or through differential DM (caused by the different electron column through the different paths across frequency).
\citet{Cordes_2017} explored the possibility of plasma lensing\footnote{Also see \citet{2020arXiv201015145L} for a discussion of possible plasma lensing in FRB\,121102, evidenced by a large spectral peak at 7.1\,GHz.} of FRBs from lenses within the host galaxy, considering 1D overdense (divergent) Gaussian lens of width $a$, extra column density $\dm_l$, and a distance between the source and lens of $d_{sl}$.  The focal length of a lens must be less than the distance to the observer from the lens for caustics to form, expressed as the constraint 
\begin{equation}
0.65 \left(\frac{d_{sl}}{\rm{pc}}\right) \left(\frac{\dm_{l}}{\rm{pc}\,\rm{cm}^{-3}}\right)  \left(\frac{a}{\rm{AU}}\right)^{-2} \left(\frac{\nu}{\rm{GHz}} \right)^{-2} \geq 1 \;\;.
\label{eq:focal}
\end{equation}
The formation of caustics depends very strongly on small-scale variations of DM,  since even a small $\dm_{l}$ can form caustics with sufficiently small $a$, due to the $a^{-2}$ dependence.  As an example of this, strong lensing in the Black Widow pulsar B1957+20 is seen to occur in regions where $\Delta \dm \sim 10^{-4}\,\mdmunits$ over ${\sim}1000\,$km  \citep{Main:2018kfc}. To make informed estimates of lensing occurring in FRB\,121102, one would like to have measurements of the smaller scale \dm\ variations.

Lensing can occur in proximity of the source of the FRB or farther out in the host galaxy. In the first case, we can rely on the measured rotation measure (RM) variations of $2200$\,rad\,m$^{-2}$ over 3 days \citep{hilmarsson+20}, coupled with an estimate of the magnetic field $B$, in order to estimate the several day variations of DM. \citet{hilmarsson+20} fit the measured RM variations with the model of \citet{margalit+18}, which assumes winds and flares from a young magnetar driving a constant-velocity expansion of a highly magnetized nebula. They try three different model conditions of the free magnetic energy of the magnetar, onset of the magnetar's active period, and radial velocity; their fit conditions and best fit values are given in Table 4 of \citet{hilmarsson+20}. From the range of outflow velocities, magnetic energies, and best-fit nebular ages, one can derive a range of expanding shell radii of $R=0.04{-}0.12$\,pc, an inferred magnetic field strength within the nebula of $B= 0.76{-}1.36$\,G, and extra DM in this region of $\dm = 0.09{-}0.16\,\mdmunits$. Using the range of magnetic fields, the inferred variation of DM over 3 days (from $\Delta\textrm{RM}\sim2200$\,rad\,m$^{-2}$ over 3 days) is $\Delta \dm \sim 0.002{-}0.0036\,\mdmunits$.

With these estimates, it is not impossible to get lensing, but to satisfy the focal constraint of Eq. \ref{eq:focal} at our observing frequencies one would need larger values of DM, or fluctuations on smaller scales. While lensing could occur, there are not sufficient electrons to create differential DMs of ${\sim}1\,\mdmunits$, and geometric time delays of lensing would be of order $\upmu$s rather than several ms to explain the precursors as bursts preceding an echo. 

If lensing is occurring in the host galaxy (i.e. if the DM in the host is de-coupled from the region causing the large RM), then there is much more material able to cause lensing. As mentioned in \citet{Cordes_2017}, lensing in the host galaxy can create caustics, which could cause geometric time delays up to ${\sim}10$\,ms, with differential DM of ${\sim} 1\,\mdmunits$. 

We confirm that lenses with $a \sim 10$\,AU satisfy the focal constraint of Eq. \ref{eq:focal} and are consistent with the measured DM variations ${\sim}1 \,\mdmunits$. 
We performed simple geometric optics simulations following the ideas presented \citet{Cordes_2017}, for an overdense or underdense Gaussian lens.  We place the lenses at a distance of $d_{sl} = 500\,$pc, with lens sizes of  as $a \sim 10$\,AU with $\Delta$DM$\,=1\,\mdmunits$ for $DM(x_{\rm lens})$, and use the unknown relative velocity between the source and lens of $v=100\,$km/s.  The phase across the lens is $\phi(x_{\rm lens}) = \phi_{\rm g}(x_{\rm lens}) + \phi_{\rm DM}(x_{\rm lens})$, where $\phi_{\rm g}$ is the geometric phase; images are defined as positions of stationary phase (ie. where $\frac{d\phi(x_{\rm lens})}{d x_{\rm lens}} = 0$), and the magnification of each image is given by $\frac{d^{2}\phi(x_{\rm lens})}{d x_{\rm lens}^{2}}$.  The results of the simulation are shown in the top panels of Figure \ref{fig:lens_sim}. Along with a magnification $\mu(f)$, each image has geometric delay $\tau_{g}(f)$ and differential DM $DM(f)$.  The total magnification $\mu(f, t)$ is computed as the incoherent sum of the images.  To qualitatively assess how the morphology of bursts could be affected by lensing, we create mock bursts (an achromatic Gaussian with 0.5\,ms width), and for a given time, apply $\mu(f)$, $\tau_{\rm g}(f)$, and the time delays associated with DM($f$) -- we show three examples of mock bursts for both a divergent and convergent lens in the bottom panels of Figure \ref{fig:lens_sim}. 

Several features comparable to the observed burst structures can be produced, including dimming at lower frequencies, and chromatic DMs (or apparent differential DMs from $\tau_{\rm g}(f)$).  Near a cusp caustic (or ``catastrophe''), the flux could sharply decrease below the focal frequency, associated with a sharp increase in DM (eg. Figure \ref{fig:lens_sim}, panel b, left), and in regions of multiple images, one can see multiple echoed copies of the burst with different apparent DM (eg. panel c, right). In this example, such features would last on ${\sim}$day--month timescales for lensing in the host galaxy, but this depends on the size and distance of the lens, the transverse velocity of the FRB, and the magnification of the caustics. However, we caution that these are highly simplified and idealistic simulations; a proper treatment would consider interference between images, and more realistic lenses.  
In geometric optics, the magnification formally diverges at caustic boundaries; in such regions, wave optics will become important \citep{grillo+2018, jow+20} and we may instead see a smooth transition of intensity across frequency (see the lower part of Figures \ref{fig:03_wfall} and \ref{fig:05_wfall}), and possibly interference effects. Interference effects could induce changes on much smaller timescales; for the above example, a very rough timescale for interference effects is $t_{\rm diff} \sim \frac{\lambda}{v_{\rm sl}} \frac{d_{\rm sl}}{2c \tau_{g}}  \sim 40$\,s for the simulated lens values, and $\lambda = 25$\,cm, $\tau_{g}=100\mu$s.

\subsubsection{Polarisation}
\label{ssec:pol}
FRB\,121102 resides in an extreme magneto-ionic environment, so lensing scenarios may be distinguished using polarisation properties. The refractive index in a magnetised plasma for left (L) and right (R) circular polarisation states is
\begin{equation}
n_{L,R} = \sqrt{1 - \frac{f_{p}^{2}}{f(f \mp f_{B,||}) }} \approx 1 - \frac{1}{2} \frac{f_{p}^{2}}{f^{2}}\left(1 \pm \frac{f_{B,||}}{f}\right),
\end{equation}
where $f_{p} \approx 9\,\mathrm{kHz} \sqrt{n_{e} / cm^{-3}}$ is the plasma frequency, and $f_{B,||} \approx 2.8\,\mathrm{MHz} \sqrt{B_{||} / G}$ is the cyclotron frequency of the parallel magnetic field.  Different refractive indices would imply a different group velocity between the two polarisation states, which results in a Faraday delay of 
\begin{equation}
\tau_{FR} \approx 0.0572\,\mathrm{ns} \left( \frac{\rm{RM}}{\mathrm{rad}\,\rm{m}^{-2}} \right) \left(\frac{f}{\mathrm{GHz}} \right)^{-3}.
\end{equation}  
At 1\,GHz, $\rm{RM}\approx10^{5}$\,rad\,m$^{-2}$ results in a delay of $\approx 5.7\upmu$s.
Additionally, if lensing effects are important, the focal frequencies between the two polarisations will differ by the cyclotron frequency \citep{li+19}, $\Delta f \approx 2.8\, \mathrm{MHz}/G$.

These effects are unlikely to be seen in incoherent filterbank data (as those presented in this paper), but could potentially be revealed by coherently comparing the timestreams between polarisations.  Searching for a coherent correlation may also reveal whether the precursors, which are qualitatively quite similar to the bursts they precede, are copies or echoes. An example of such techniques is shown in \citet{main+17}, who coherently correlate nearby giant bursts in PSR B1957+20.  In addition, Faraday delays will be much more evident at lower frequencies, scaling as $\nu^{-3}$.  In such an environment, it may be possible to observe higher order effects such as Faraday conversion \citep{vedantham+19, li+19}. However, at lower frequencies the source is likely depolarized within individual channels, and it may only be possible to detect these effects coherently, using voltage data of bursts.

\subsection{Dimming below 1250\,MHz}

We do not expect that the observed dimming below 1250\,MHz is caused by absorption, unless absorption is highly variable, since a burst was detected by CHIME \citep{Josephy_2019} at much lower frequencies.  However, absorption could still play a role if  conditions are changing along the line-of-sight.  In fact, if FRB\,121102 is in an orbit \citep[a possibility to explain the periodicity of its active period;][]{rajwade+20}, absorption and lensing could easily be phase dependent. This may be assessed by looking for a phase dependence of burst properties. 

\subsection{Frequency drifts}

A common feature of repeating FRBs is a downward drift in frequency \citep[e.g.][]{andersen_2019,Fonseca_2020,Hessels2019}. In Paper I, the structure-optimised DMs reported here were used to characterise the drift-rates of bursts 07 and 11 using a 2D ACF method. The structure-optimised DM for burst 07 has since been updated, and thus we present the revised value for the frequency drift. Previously, burst 07 was sub-banded and the drift rate was measured at a center frequency of 1400\,MHz over a bandwidth of 214\,MHz. Here, we additionally measure the drift rate at a center frequency of 1284\,MHz over 856\,MHz. The drift rate of burst 08 is also measured at a center frequency of 1284\,MHz over 856\,MHz. Results are presented in Table \ref{tab:drift}.

\begin{table}
	\centering
	\caption{Measured drift rates for various sub-bands of bursts 07, 08 and 11.}
	\label{tab:drift}
	\begin{tabular}{ l c c c}
        \hline\hline
        Burst & Center frequency & Drift rate & Bandwidth \\
        & (MHz) & (MHz\,ms$^{-1}$) & (MHz)  \\
        \hline
        07 & 1284 & $-102.2\pm4.1$ & 856 \\ 
           & 1400 & $-26.8\pm0.7$ & 214 \\ 
        \hline
        08 & 1284 & $-111.8\pm1.6$ & 856 \\ 
        \hline
        11 & 906 & $-27.5\pm0.5$ & 100 \\ 
           & 1128 & $-85.0\pm1.8$ & 544 \\ 
           & 1284 & $-53.7\pm1.1$ & 856 \\ 
           & 1400 & $-16.5\pm0.2$ & 214 \\ 
        \hline
    \end{tabular}
\end{table}

These are consistent with those published between $600{-}6500$\,MHz, with a slope of $\alpha=-0.147\pm 0.014\, \text{ms}^{-1}$. No upward drift is reported. Where bursts consist of two sub-bursts (e.g. bursts 03 and 05), however, it is unclear whether the bursts are independent or occur within the same burst envelope. In the latter case, burst 03 may be an example of upward drifting: the second panel of Figure \ref{fig:gallery} shows a faint precursor burst between ${\sim}$1000--1200\,MHz, followed by a main burst between ${\sim}$1000--1700\,MHz. The precursor, however, may just be intrinsically fainter overall than the second component or fainter at higher frequencies. Similar potential upward drifting behaviour has been reported before in repeating (periodic) FRB\,180916.J0158+65 with a burst separation of ${\sim}60$\,ms \citep{Amiri_2020,Chawla_2020}, in the apparently one-off FRB\,190611 with a burst separation of ${\sim}0.7$\,ms \citep{2020MNRAS.497.3335D} and in the Galactic FRB\,200428 with a burst separation of ${\sim}29$\,ms \citep{Bochenek_2020,Andersen_2020}.

%%%%%%%%%%%%%%%%%%%%%%%%%%%%%%%%%%%%%%%%%%%%%%%%%%
\section{Conclusion}
\label{sec:conclude}
In this paper, we calculated the structure-optimised DMs for 10 out of the 11 FRB\,121102 bursts detected by the MeerKAT radio telescope originally presented in \citet{2020MNRAS.496.4565C}. Two independent methods were used to do so -- ACFs \citep[following ][]{Lange1998} and \code{DM\_phase} \citep{Seymour_2019}. We find that while results largely agree, care should be taken when selecting an ``optimal'' DM: where results are ambiguous, it is not always clear which burst profile best represents the burst at origin. Potentially unresolved sub-components further complicate accurate DM determination. The main bursts of 03 and 05 illustrate this point well: while they appear to deviate from the standard $t\sim\nu^{-2}$ relationship when de-dispersed to their structure-optimised DMs, it is possible that the bursts are actually composed of unresolved downward drifting sub-bursts.

If we consider the main bursts of 03 and 05 (without considering any unresolved components) we find that at lower frequencies the DMs are ${\sim}1{-}2$\,\dmunits\ lower than at higher frequencies. This may imply lensing, which we show can plausibly account for such differences if the lensing occurs in the host galaxy. In such a scenario, no single DM can describe the intrinsic burst morphology.

Two of the reported bursts have precursors (bursts 03 and 05). The time difference between bursts is comparable to the longest duration burst reported for FRB\,121102 \citep[${\sim}39$\,ms; ][]{2020arXiv200803461C}, and thus the two bursts may plausibly result from one event. Polarisation information and RMs are unfortunately unavailable. If it were, a coherent correlation between polarisations could reveal whether burst pairs are actually echoes \citep[e.g.][]{main+17}.

Determining the average DM for the epoch is not elementary. Potentially unresolved sub-components of the bursts would greatly influence the measured DM. As such, we found $\langle\dm\rangle = 563.5\pm 0.2 (\text{sys}) \pm 0.8 (\text{stat})\,\mdmunits$ by considering only bursts whose DM could be established with reasonable reliability. This is consistent with measurements of FRB\,121102 by \citet{Josephy_2019} and \citet{Oostrum_2020} taken one year prior to our sample. Interpolating using these results, as well as 2016 measurements by \citet{Hessels2019}, we obtain a mean increase of 1\,\dmunits\ per year. Future observations will help establish whether the increase is indeed persistent.

To establish whether the potential deviation from the $t\sim\nu^{-2}$ relationship observed in the main bursts of 03 and 05 is plausible, one might compare the individual structure-optimised DMs to the average DM of the epoch. At a 1$\sigma$ confidence level, the DM of burst 05 ($564.5\pm0.3\,\mdmunits$ and $564.4\pm0.3\,\mdmunits$ for ACF and \code{DM\_phase}, respectively) is consistent with the average. The structure-optimised DM of burst 03, however, is not. One would require a deviation of ${\sim}1{-}2\,\mdmunits$ from the average DM of the epoch. Such is possible (note that bursts 03 and 05 have different \dmstruct\ values for near-identical structures), but the more mundane option may be more likely: that the bursts do not exhibit a deviation from the expected cold plasma dispersion relationship but instead have downward drifting sub-components that are not resolved.

Many of the issues highlighted in this paper are seen in our wide band, and thus some narrow band data may not have been sensitive to these effects. It will be interesting going forward to see if similar behaviour is observed in higher resolution, wide bandwidth data.

%%%%%%%%%%%%%%%%%%%%%%%%%%%%%%%%%%%%%%%%%%%%%%%%%%
\section*{Acknowledgements}
The authors would like to thank the anonymous referee for their invaluable feedback and perspective. The authors would like to thank Daniele Michilli and Andrew Seymour for help with \code{DM\_phase}, and thank Fang Xi Lin for help and advice with lensing simulations. The authors would also like to thank SARAO for the approval of the DDT MeerKAT request and the CAM/CBF and operator teams for their time and effort invested in the observations. The MeerKAT telescope is operated by the South African Radio Astronomy Observatory (SARAO), which is a facility of the National Research Foundation, an agency of the Department of Science and Innovation.
EP is supported by a L'Or\'{e}al-UNESCO For Women in Science Young Talents Fellowship and by a PhD fellowship from the South African National Institute for Theoretical Physics (NITheP). AW acknowledges funding from the South African Research Chairs Initiative funded by the National Research Foundation and Department of Science and Technology. MC, BWS, FJ, KR, VM and MCB acknowledge funding from the European Research Council (ERC) under the European Union's Horizon 2020 research and innovation programme (grant agreement No 694745). 

\section{Data Availability}
The data underlying this article will be shared on reasonable request to the corresponding author.

%%%%%%%%%%%%%%%%%%%%%%%%%%%%%%%%%%%%%%%%%%%%%%%%%%
\bibliographystyle{mnras}
\bibliography{DM_struct.bib}

\bsp
\label{lastpage}

\end{document}